\documentclass[twocolumn,  tighten,  twocolappendix]{aastex631} 

\shorttitle{Warped Nature of the Fundamental Plane of Early-type Galaxies}
\shortauthors{Yoon \& Park (2022)}

\begin{document}

\title{The Fundamental Plane Is Not a Plane: Warped Nature of the Fundamental Plane of Early-type Galaxies and Its Implication for Galaxy Formation}

\email{yyoon@kias.re.kr}

\author[0000-0003-0134-8968]{Yongmin Yoon}
\affiliation{School of Physics, Korea Institute for Advanced Study (KIAS), 85 Hoegiro, Dongdaemun-gu, Seoul, 02455, Republic of Korea}
\affiliation{Korea Astronomy and Space Science Institute (KASI), 776 Daedeokdae-ro, Yuseong-gu, Daejeon, 34055, Republic of Korea}

\author[0000-0001-9521-6397]{Changbom Park}
\affiliation{School of Physics, Korea Institute for Advanced Study (KIAS), 85 Hoegiro, Dongdaemun-gu, Seoul, 02455, Republic of Korea}

\begin{abstract}
Based on $16,283$ early-type galaxies (ETGs) in $0.025\le z_\mathrm{spec}<0.055$ from Sloan Digital Sky Survey data, we show that the fundamental plane (FP) of ETGs is not a plane in the strict sense but is a curved surface with a twisted shape whose orthogonal direction to the surface is shifted as the central velocity dispersion ($\sigma_0$) or mean surface brightness within the half-light radius ($\mu_e$) changes. When ETGs are divided into subsamples according to $\sigma_0$, the coefficient of $\mu_e$ of the FP increases, whereas the zero-point of the FP decreases at higher $\sigma_0$. Taking the $z$ band as an example, the coefficient of $\mu_e$ rises from $0.28$ to $0.36$ as $\sigma_0$ increases from $\sim100$ to $\sim300$ km s$^{-1}$. At the same time, the zero-point of the FP falls from $-7.5$ to $-9.0$ in the same $\sigma_0$ range. The consistent picture on the curved nature of the FP is also reached by inspecting changes in the FP coefficients for ETG subsamples with different $\mu_e$. By examining scaling relations that are projections of the FP, we suggest that the warped nature of the FP may originate from dry merger effects that are imprinted more prominently in ETGs with higher masses.
\end{abstract}
\keywords{Early-type galaxies (429) --- Galaxy evolution (594) ---  Galaxy formation (595) --- Galaxy mergers (608) --- Galaxy properties (615) --- Red sequence galaxies (1373)}

\section{Introduction}\label{sec:intro}

The old stellar populations and low star formation activities due to depletion of cold gas are the typical properties of early-type galaxies (ETGs). For this reason, the majority of bright ETGs have red colors ($u-r\gtrsim2.7$) in optical bands \citep{PC2005,Gallazzi2006,Choi2007,Choi2010,Schawinski2014,YP2020}. In addition, ETGs have centrally concentrated light distributions and relatively simple and smooth shapes compared with late-type galaxies \citep{PC2005,Choi2010,Nair2010}. In the kinematic sense, ETGs are virialized systems that are expected to satisfy the balance between potential and kinetic energy such as
\begin{equation}
\sigma^2 \propto  \frac{M_\mathrm{dyn}}{R} \propto \frac{M_\mathrm{dyn}}{L}IR, 
\label{eq:bal}
\end{equation} 
in which $\sigma$ is the velocity dispersion of a galaxy, $M_\mathrm{dyn}/L$ is the dynamical mass-to-light ratio, $R$ is the galaxy size, and $I$ is the galaxy surface brightness ($I\propto L/R^2$). According to this condition, the three observational quantities of ETGs, which are the logarithm values\footnote{The base-10 logarithm is used for all the logarithm values in this paper.} of the half-light radius ($R_e$), the central velocity dispersion ($\sigma_0$), and the mean surface brightness within $R_e$ ($\mu_e=-2.5\log I_e$), form a plane in the three-dimensional parameter space, known as the fundamental plane (FP; \citealt{Djorgovski1987,Dressler1987,YP2020}), described by
\begin{equation}
\log R_e=a\log\sigma_0+b\mu_e+c.
\label{eq:fp}
\end{equation}

Studies based on observational data have shown that the coefficients of the FP ($a\sim1$--$1.5$ and $b\sim0.3$; \citealt{Bernardi2003b,Jun2008,Hyde2009b,LaBarbera2010,Cappellari2013,Saulder2013,YP2020}) are smaller than the values expected for fully virialized ETGs with constant $M_\mathrm{dyn}/L$ (i.e., $a=2$ and $b=0.4$; see Equation \ref{eq:bal}). There have been a few studies suggesting, based on cosmological simulations, that different degrees of dissipational merger effects imprinted in ETGs with different masses can cause smaller FP coefficients than expected \citep{Dekel2006,Robertson2006,Hopkins2008}. In this way, the coefficients of the FP give useful information for understanding the formation and evolution of ETGs \citep{Bertin2002,Trujillo2004,Cappellari2006,Dekel2006,Robertson2006,Hopkins2008}.

ETGs have diverse formation histories contrary to their smooth and uncomplicated shapes. For instance, massive ETGs with large sizes grow by multiple dry mergers \citep{Lauer2007,Bernardi2011b,Oogi2013,Oogi2016,Yoon2017}, while ETGs with compact light distributions are likely to be formed from gas-rich mergers \citep{Mihos1994,Robertson2006,Hopkins2008,Kormendy2009,YL2020}. The nonuniform formation mechanisms cause ETGs to follow curved/nonlinear scaling relations between ETG properties. For example, the scaling relations between luminosities (or stellar masses) and velocity dispersions/sizes/surface brightness are nonlinear \citep{Choi2007,Desroches2007,Hyde2009a,Bernardi2011b,Kormendy2013,Yoon2017} in the sense that very luminous ETGs have lower velocity dispersions, larger sizes, and fainter surface brightness than expected values that are extrapolated from the scaling relations based on less luminous ETGs. Similarly, \citet{Bernardi2007} and \citet{Samir2020} showed that the scaling relations for brightest cluster galaxies (BCGs) are different from those for less luminous normal ellipticals.

The FP is the basis of various scaling relations of ETGs, since scaling relations mentioned above are projections or variations of the FP. Several previous studies found possible hints for curvature in the FP \citep{Jorgensen1996,Zaritsky2006,D'Onofrio2008,Hyde2009b}. For example, \citet{Jorgensen1996} found a possible hint that the FP is slightly curved for highly luminous galaxies. \citet{Zaritsky2006} showed a systematic decline in the coefficient of the FP as the galaxy velocity dispersion changes. \citet{Hyde2009b} suggested that the FP is probably warped at the low-mass/small-size end of ETGs. Similarly, \citet{D'Onofrio2008} speculated that the FP is a curved surface based on their data. However, some studies showed that the curved nature can be an artifact from a selection bias in the geometry of the ETG distributions \citep{Bernardi2003b,D'Onofrio2008,Gargiulo2009,Hyde2009b,Nigoche-Netro2009}.

Here we intensively examine the FP with a large number of galaxies to definitely determine the warped/nonlinear nature of the FP. By doing so, we discover that the FP is actually a slightly curved and twisted surface.

This paper is organized as follows. The galaxy sample is described in Section \ref{sec:sample}. The plane- and line-fitting methods are shown in Section \ref{sec:fit}. We describe binning of the ETG sample and possible biases from the fitting in Section \ref{sec:binning}. The result of this study and its implication for the formation of ETGs are described in Sections \ref{sec:results} and \ref{sec:imp}, respectively. Section \ref{sec:comp} shows comparisons between this study and previous ones conducted so far. Finally, we summarize our study in Section \ref{sec:summary}. Throughout this study, we use the AB magnitude system. The cosmological parameters used here are $H_0=70$ km s$^{-1}$ Mpc$^{-1}$, $\Omega_{\Lambda}=0.7$, and $\Omega_m=0.3$.
\\

\section{Sample}\label{sec:sample}
Our sample used in this study is identical to that of \citet{YP2020}. We used galaxies that have spectroscopic redshifts from the Sloan Digital Sky Survey (SDSS) and that are classified as ETGs in the KIAS value-added catalog \citep{Choi2010}. This catalog, which is based on SDSS Data Release 7 \citep[DR7;][]{Abazajian2009}, classified galaxies into early types or late types using three parameters: $u - r$ color, $g-i$ color gradient, and inverse concentration index in the $i$ band. Details about the morphology classification are in \citet{PC2005} and \citet{Choi2007}. In brief, ETGs are classified by the following criteria: (1) high concentrated light distributions, and (2) red $u-r$ color with slightly negative (bluer outside) or flat $g-i$ color gradient as is the case for most ETGs, or blue $u-r$ color with positive $g-i$ color gradient to include blue ETGs. We note that the reliability and completeness of this classification are almost $90\%$. The visual inspection was also performed to improve the automatic classification, thereby rectifying $7\%$ of the inspected galaxies.

To cross-check the morphologies of our final ETG sample, we examined the weight (from 0 to 1) of the de Vaucouleurs fit component in the combined model of the de Vaucouleurs fit and the exponential disk fit (the parameter $fracDev$ in SDSS data) in the $r$ band. By doing so, we found that $70\%$ of ETGs in the final sample have $fracDev>0.95$, while $94\%$ of ETGs have $fracDev>0.71$.\footnote{For the $z$ band, $70\%$ of ETGs have $fracDev>0.99$, while $94\%$ of ETGs have $fracDev>0.77$.} This indicates that the de Vaucouleurs model is a better description for ETGs used in this study than the exponential model. However, it is still possible that neither of them is a proper model.

We note that the conclusions of this study are not sensitive to the definition of ETGs, since use of different criteria to define ETGs (e.g., criteria used in \citealt{Saulder2013}) does not change our main results.

We used ETGs within the redshift range of $0.025\le z_\mathrm{spec}<0.055$. The lower limit of 0.025, which corresponds to the distance of $\sim100\,\mathrm{Mpc}$, was set for the purpose of mitigating the peculiar velocity effects that can distort distance-dependent galaxy properties at very low redshifts. 

The absolute magnitudes were derived by 
\begin{equation}
M=m -\mathrm{DM}-K+Qz_\mathrm{spec}, 
\label{eq:abm}
\end{equation}
where $m$ is the galactic-extinction-corrected apparent magnitude, DM indicates the distance modulus, and $K$ is the $k$-correction. $Q$ is the parameter for evolution correction, so that $Qz_\mathrm{spec}$ is a term to correct for the passive evolution of galaxy luminosity. For $m$, we used model magnitudes from the two-dimensional (2D) de Vaucouleurs fits. The galactic extinction corrections were applied to the magnitudes using the dust maps of \citet{Schlegel1998}. The $k$-correction values were computed using the IDL software of \citet{Blanton2007}, which calculates $k$-correction values from spectral energy distribution (SED) models fitted to photometric magnitudes of the five SDSS bands. The SED models are based on the \citet{Chabrier2003} initial mass function (IMF) and \citet{Bruzual2003} stellar population models\footnote{\url{http://www.bruzual.org/bc03/}} with various ages and metallicities.

The correction for passive evolution of galaxy luminosity is negligible in this study, since our ETGs are at low redshifts. Therefore, we defined $Q$ in a very simple way that uses simple stellar populations (SSPs) of \citet{Bruzual2003} models with a \citet{Chabrier2003} IMF. We calculated the variations in luminosities during 2.4 Gyr owing to the passive evolution in SSPs with ages of $10$--$13$ Gyr and metallicities of $Z=0.008$--$0.05$. By doing so, we found that the passive evolutions in magnitudes of the $g$, $r$, $i$, and $z$ bands, on average, correspond to $Q$ values of 1.26, 1.13, 1.07, and 1.02, respectively.\footnote{These $Q$ values are slightly larger than those used in \citet{Bernardi2003b}.} 

In this study, we used ETGs with $M_r\le-19.5$. We note that $M_r=-19.5$ corresponds to the $r$-band magnitude limit for the spectroscopic target selection ($m_r\approx17.77$) at the upper redshift limit of $z_\mathrm{spec}=0.055$. The number of ETGs in the volume-limited sample of $M_r\le-19.5$ and $0.025\le z_\mathrm{spec}<0.055$ is 22,474.

For physical sizes of ETGs, we used half-light radii (or effective radii) of the 2D de Vaucouleurs models. The half-light radius and the semi-major axis length $a_\mathrm{deV}$ are connected by
\begin{equation}
R_e=a_\mathrm{deV}\sqrt{q},
\label{eq:size}
\end{equation}
in which $q$ is the axis ratio of the 2D de Vaucouleurs model. In this study, we used the ETGs whose $r$-band $q$ values are larger than or equal to $0.3$ to exclude edge-on galaxies. By this $q$ cut, 1196 galaxies are excluded. The number of remaining ETGs is 21,278. 

The estimated stellar velocity dispersions within the aperture of the SDSS fiber were converted to central velocity dispersions within one-eighth of the half-light radius by applying the correction equation in \citet{Jorgensen1995}: 
\begin{equation}
\sigma_0=\sigma_\mathrm{est}\left(\frac{r_\mathrm{fiber}}{r_\mathrm{ang}/8}\right)^{0.04}, 
\label{eq:vel}
\end{equation}
where $\sigma_\mathrm{est}$ is the estimated velocity dispersion, $r_\mathrm{fiber}$ is the radius of fibers ($r_\mathrm{fiber}=1.5\arcsec$), and $r_{\mathrm{ang}}$ is the angular half-light radius in arcseconds. 

The instrumental dispersion (spectroscopic sampling) of the SDSS spectrograph is 69 km s$^{-1}$ per pixel, and the resolution of SDSS galaxy spectra calculated from the autocorrelation of stellar template spectra is $\sim90$ km s$^{-1}$ \citep{Bernardi2003a}. Thus, previous studies suggested that low stellar velocity dispersions less than $\sim90$ -- $100$ km s$^{-1}$ are not reliable \citep{Bernardi2003a,Saulder2013}. Hence, ETGs with $100$ km s$^{-1} \le\sigma_0<420$ km s$^{-1}$ were used in this study.\footnote{The use of $\sigma_\mathrm{est}$ instead of $\sigma_0$ for the velocity dispersion cut does not change our conclusion of this study, since the degree of the aperture correction by Equation \ref{eq:vel} is not significant ($\sim9\pm4$ km s$^{-1}$). Here we applied the $\sigma_0$ cut to the sample for consistency with our previous study on the FP \citep{YP2020}.} The upper limit was set to be $420$ km s$^{-1}$. This is because SDSS used template spectra convolved to the maximum velocity dispersion of $420$ km s$^{-1}$ for the velocity dispersion measurements, so that use of velocity dispersions larger than $420$ km s$^{-1}$ is not recommended. Applying the $\sigma_0$ cut, the number of ETGs is 16,793.

Parameter $\mu_e$ is derived by 
\begin{equation}
\mu_e=m^c  + 2.5\log(2\pi r_\mathrm{ang}^2) -2.5\log(1+z_\mathrm{spec})^3,
\label{eq:mu}
\end{equation}
where $m^c =m - K + Qz_\mathrm{spec}$. The last term is for the correction of the cosmological dimming of surface brightness in the AB magnitude system.

We note that the magnitude in each band, $a_\mathrm{deV}$, $q$, and $\sigma_\mathrm{est}$ are from the photometry and spectroscopy catalogs (PhotObjAll and SpecObjAll) of DR15 \citep{Aguado2019}. The main conclusions of this study are essentially unchanged even if we make use of magnitudes and $R_e$ based on Petrosian fluxes instead of those from the de Vaucouleurs models. Throughout this study, the units of $R_e$, $\sigma_0$, and $\mu_e$ are kpc, km s$^{-1}$, and mag arcsec$^{-2}$, respectively. 

\begin{figure}
\includegraphics[width=\linewidth]{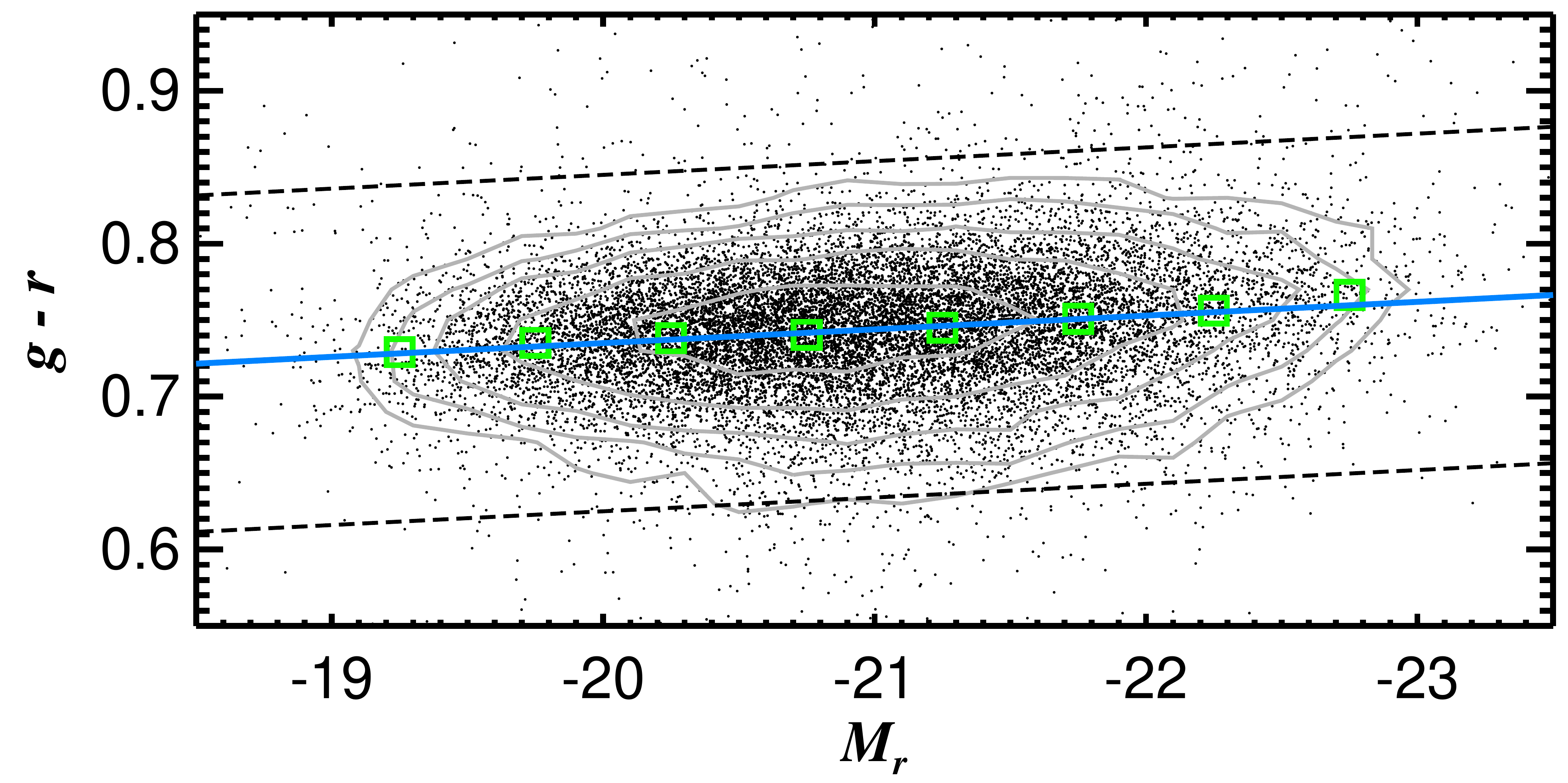}  
\centering
\caption{Color--magnitude diagram ($g-r$ vs. $M_r$) for ETGs used in this study. The black dots denote ETGs, while the green square represents the average $g-r$ value after $3\sigma$ clipping in each magnitude bin. The blue solid line indicates a line derived by the iterative least-squares fit with clipping outliers. The dashed lines are upper and lower $3\sigma$ lines ($\sigma=0.0367$), in which $\sigma$ is the standard deviation from the fitted line. The levels of the contour lines (the gray lines) represent 16, 32, 64, 128, and 256 galaxies in 2D bins whose sizes in the $x$- and $y$-axes are 0.2 and 0.02, respectively. We note that ETGs with $M_r>-19.5$ are included in this figure.
\label{fig:cmd}}
\end{figure} 

\begin{figure}
\includegraphics[width=\linewidth]{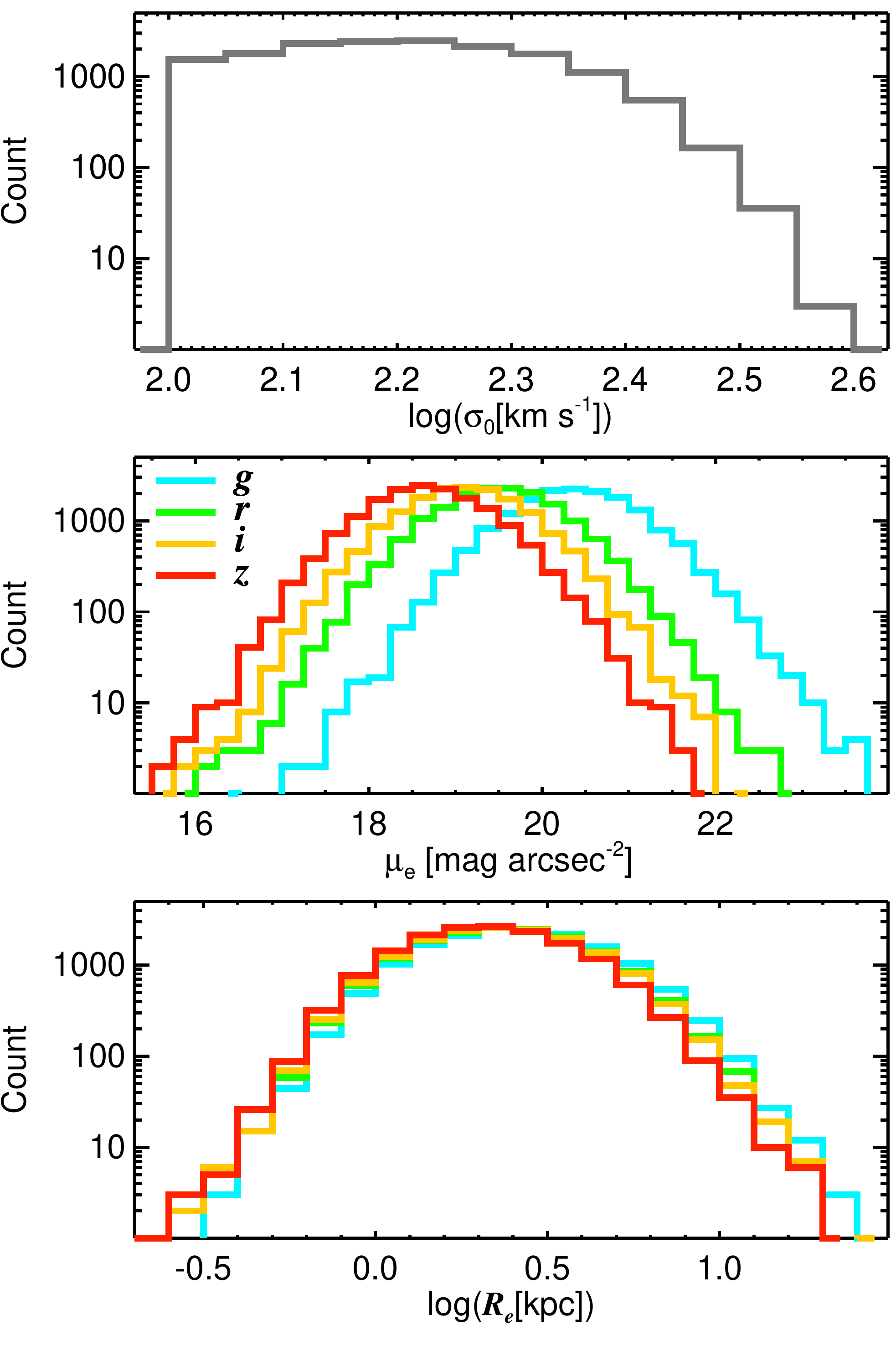}  
\centering
\caption{Distributions of $\log\sigma_0$, $\mu_e$, and $\log R_e$ for the ETG sample used in this study. The results for different bands are displayed in different colors. The bin sizes of the histograms are 0.05 dex, 0.25 mag arcsec$^{-2}$, and 0.1 dex for $\log\sigma_0$, $\mu_e$, and $\log R_e$, respectively.
\label{fig:sigdist}}
\end{figure}

Figure \ref{fig:cmd} shows the color--magnitude diagram ($g-r$ color versus $M_r$) for the ETG sample used in this study.\footnote{The $k$- and evolution corrections were applied to the $g-r$ color values.} Our ETGs form a tight red sequence, so that $g-r$ color values are distributed within the narrow range of $\sim0.2$. We performed a line fitting to the ETGs, using the least-squares method. The fitting was conducted iteratively, excluding $3\sigma$ outliers from the line until the number of galaxies converges. The equation of the fitted line (the blue line in Figure \ref{fig:cmd}) is
\begin{equation}
(g-r)=0.555-0.00899M_r.
\label{eq:cmd}
\end{equation}
The standard deviation ($\sigma$) of $g-r$ from the line is 0.0367. In this study, we used ETGs whose $g-r$ color values do not deviate more than $3\sigma$ (0.110) from the line, in order to define an uniform ETG sample in terms of optical color. Excluding a small number (510) of galaxies by this cut, the total number of ETGs in the final sample is 16,283. See Figure 3 in \citet{YP2020} for color images of typical ETGs used in this study. In Figure \ref{fig:sigdist}, we show the distributions of $\log\sigma_0$, $\mu_e$, and $\log R_e$ for the final ETG sample used in this study.
\\

\begin{figure}
\includegraphics[width=\linewidth]{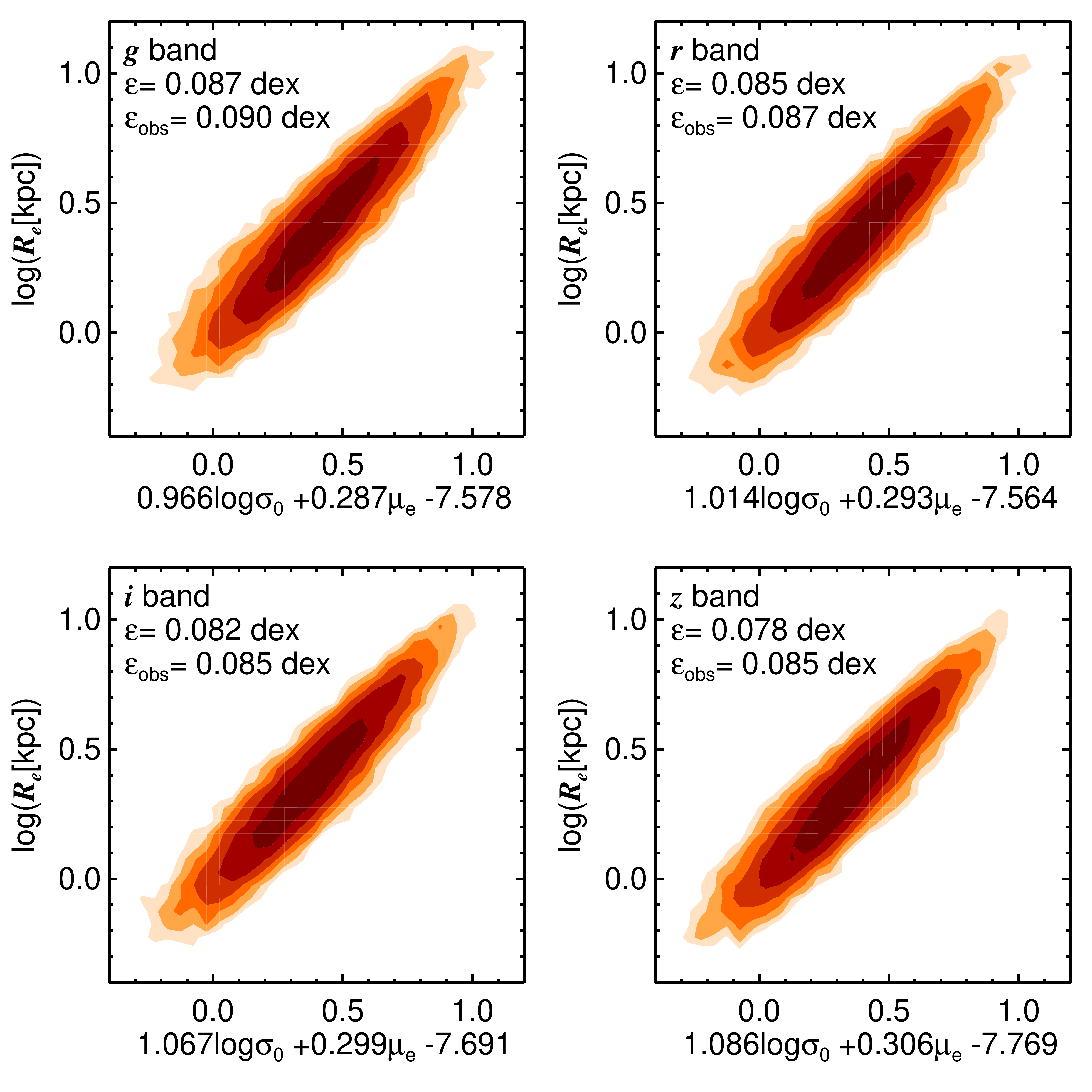}  
\centering
\caption{Edge-on view of the FP of all ETGs in the four bands ($g$, $r$, $i$, and $z$). The intrinsic scatter ($\varepsilon$) and the observed scatter ($\varepsilon_\mathrm{obs}$) in the direction of $\log R_e$ are shown in each panel. The levels of the contours indicate 6, 12, 24, 48, 96, and 192 galaxies in 2D bins whose sizes are 0.05 in the $x$- and $y$-axes.
\label{fig:FPall}}
\end{figure}

\section{Fundamental Plane Fitting}\label{sec:fit}

To fit a plane to the distribution of ETGs, we used the plane-fitting code LTS\_PLANEFIT\footnote{\url{https://www-astro.physics.ox.ac.uk/~mxc/software/}} \citep{Cappellari2013}. We briefly describe the fitting process of this code, which is basically the $\chi^2$ minimization after trimming outliers, as follows. 
\begin{enumerate}
\item Finding a subset with $h$ data points that has the smallest $\chi^2$. Here $h$ is $(N+p+1)/2$, in which $N$ is the number of total points and $p$ is the data dimension (for the plane fit, $p=3$). The $\chi^2$ is defined as
\begin{equation}
\chi^2=\sum_{i=1}^{n} \frac{(ax_i+by_i+c-z_i)^2}{(a\Delta x_i)^2+(b\Delta y_i)^2+(\Delta z_i)^2+\varepsilon^2},
\label{eq:chi}
\end{equation}
where $x=\log \sigma_0$, $y=\mu_e$, and $z=\log R_e$ for finding the FP, and $\Delta x$, $\Delta y$ or $\Delta z$ means the error of the quantity; $\varepsilon$ is the intrinsic scatter in the direction of the $z$-axis.
\item Calculating the standard deviation value of the residuals for the $h$ data points. Then, among all the $N$ data points, those that deviate more than $3\sigma$ from the fitted plane are excluded.
\item Iterating the above steps until the set of selected data points does not change further.
\item Computing the $\chi^2$ for the final data points.
\item Repeating all the above steps, varying $\varepsilon$ until $\chi^2=\nu$, in which $\nu$ is the degree of freedom.
\end{enumerate}
In this way, we determined the coefficients of the FP and the intrinsic scatter in the direction of $\log R_e$. Figure \ref{fig:FPall} shows the edge-on view of the FP of all ETGs in the four SDSS bands ($g$, $r$, $i$, and $z$). We note that this plane-fitting method is highly consistent with the maximum likelihood Gaussian fitting method that minimizes the residuals in the direction of $\log R_e$. See Appendix \ref{Appendixcomp} for comparisons between the fitting results from the two different methods.

 This fitting process is also applicable to the line fitting (e.g., $y=\alpha+\beta x$) if the $\chi^2$ is defined as
\begin{equation}
\chi^2=\sum_{i=1}^{n} \frac{(\alpha+\beta x_i - y_i)^2}{(\beta\Delta x_i)^2+(\Delta y_i)^2+\varepsilon^2}.
\label{eq:chi2}
\end{equation}
The fitting process described above clips data points from the inside out, which makes the fitting result less affected by outliers, compared with the standard $\sigma$ clipping method \citep{Cappellari2013}. 
\\

\begin{figure*}
\includegraphics[width=\linewidth]{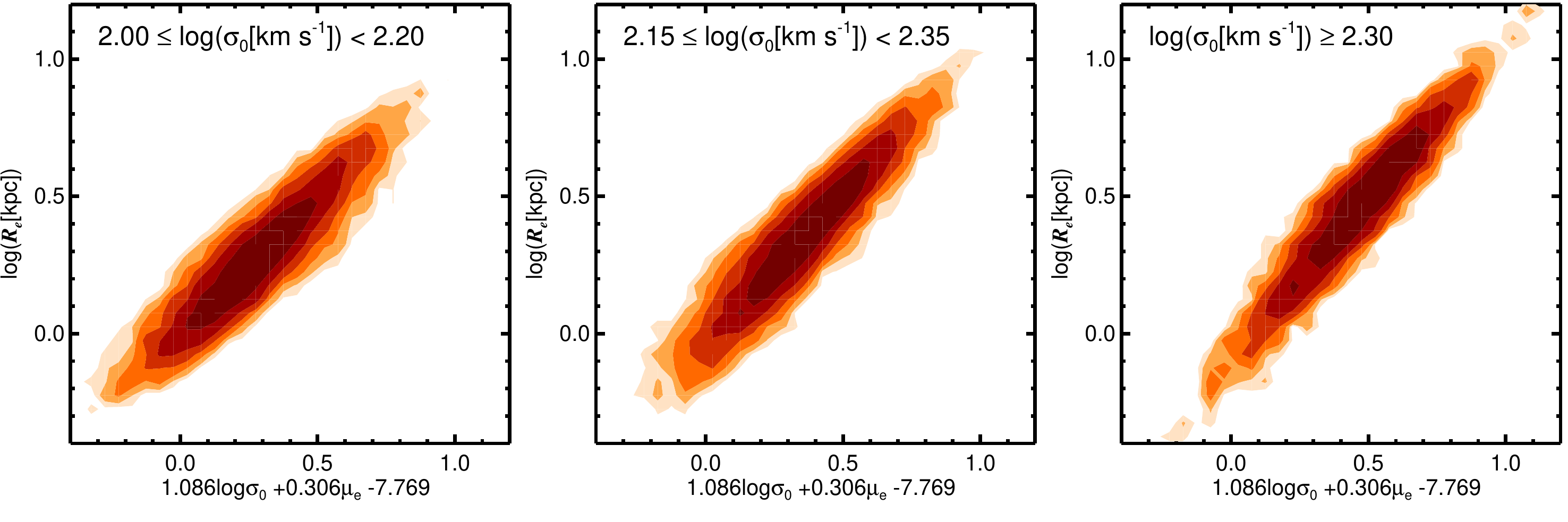}  
\centering
\caption{ETG distributions in the three-dimensional parameter space of $\log R_e$, $\log\sigma_0$, and $\mu_e$ for three different $\log\sigma_0$ bins. The parameters used here are based on the $z$ band. The ETG distributions are projected toward the edge-on view direction of the $z$-band FP of all ETGs shown in Figure \ref{fig:FPall}. The levels of the contours in each panel are scaled down from the contour levels in Figure \ref{fig:FPall} by a ratio between the number of ETGs in each $\log\sigma_0$ bin and the number of the whole ETGs. 
\label{fig:sigbin}}
\end{figure*}

\begin{figure}
\includegraphics[width=\linewidth]{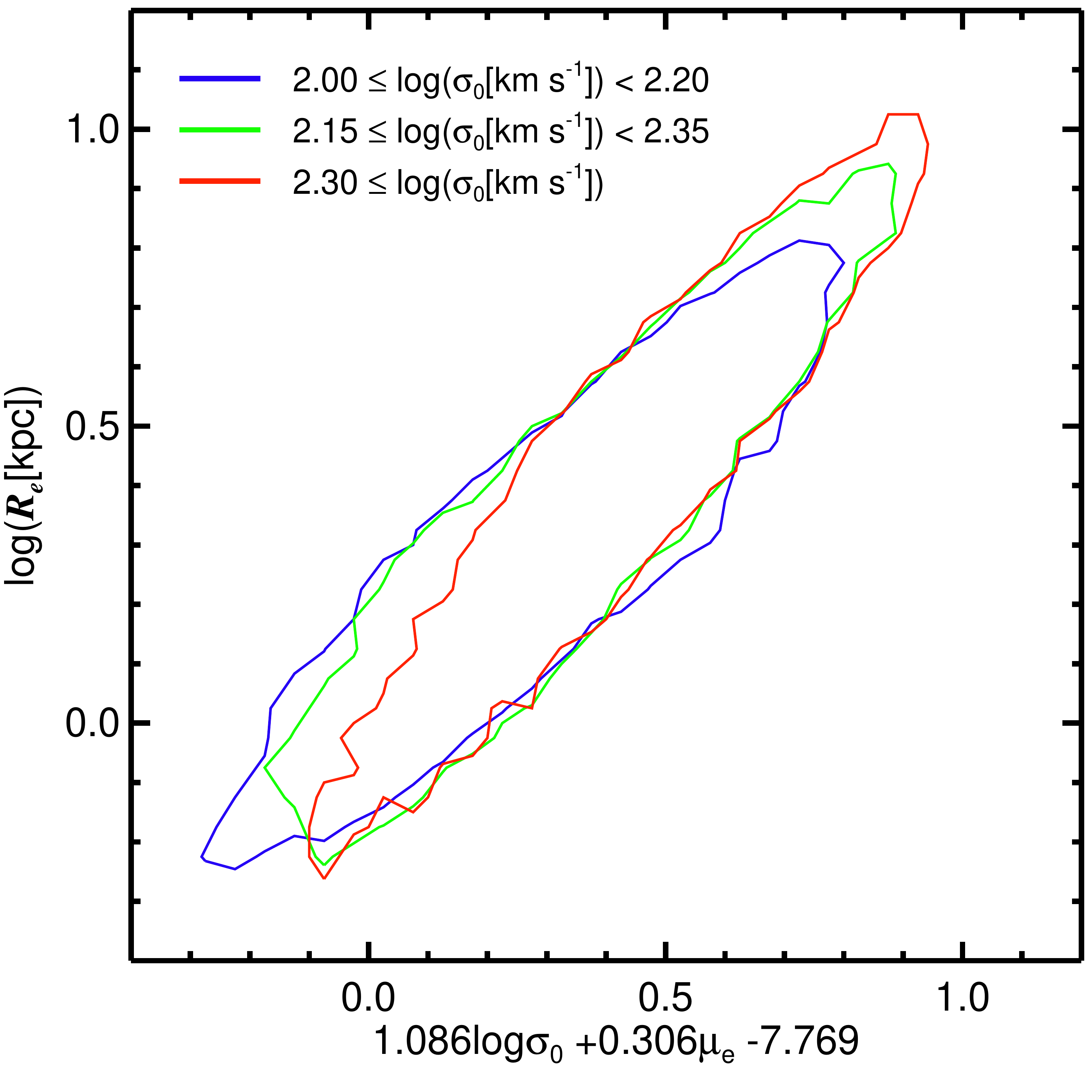}  
\centering
\caption{Superposed are equal density contours for distributions of ETGs in the three $\log\sigma_0$ bins that are projected toward the edge-on view direction of the $z$-band FP of all ETGs. The parameters used in this figure are based on the $z$ band. The contour lines indicate eight (blue), nine (green), and four (red) galaxies in 2D bins whose sizes are 0.05 in the $x$- and $y$-axes (each value roughly corresponds to one-thousandth of the number of ETGs in each $\log\sigma_0$ bin: 8010, 8805, and 3642, respectively). 
\label{fig:comp}}
\end{figure} 

\begin{figure*}
\includegraphics[scale=0.30]{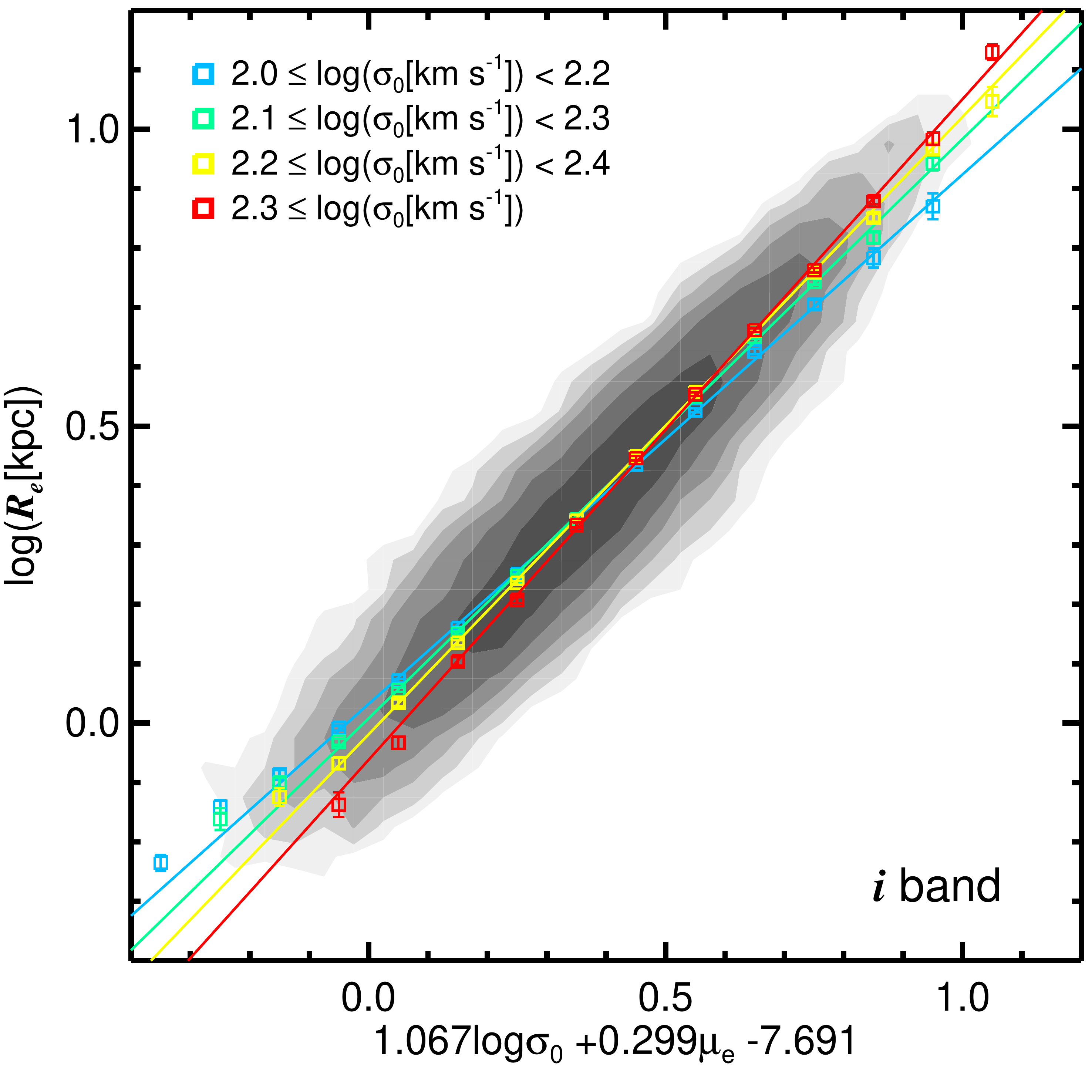}\includegraphics[scale=0.30]{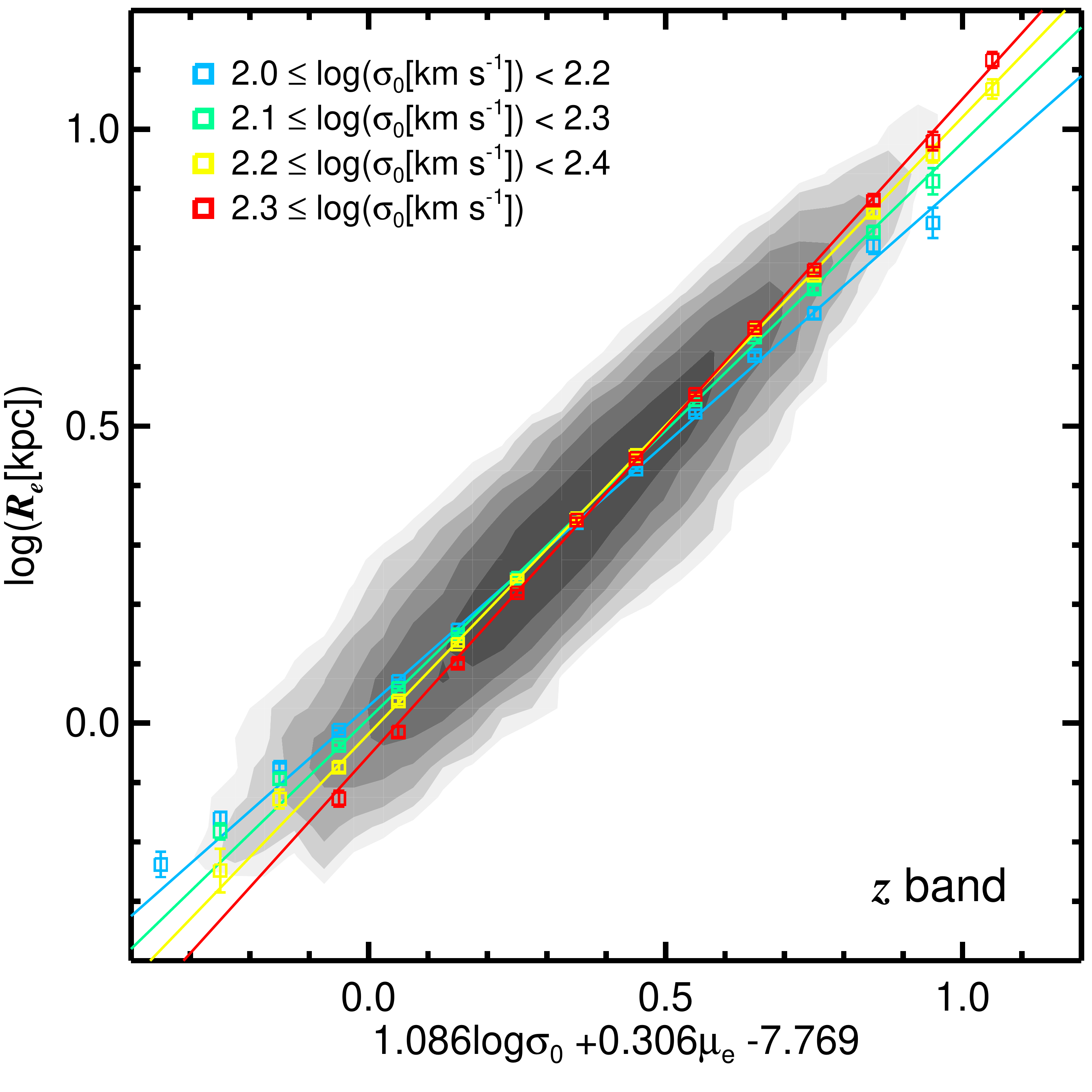}  
\centering
\caption{Slopes and zero-points of distributions of ETGs with different $\log\sigma_0$ when the distribution of the whole ETGs is projected toward the edge-on view direction of the $i$- or $z$-band FP of all ETGs. The contours represent distributions of all ETGs as shown in Figure \ref{fig:FPall}. The left panel is for the $i$-band FP, while the right panel is for the $z$-band FP. Each square indicates the mean $\log R_e$ after 3$\sigma$ clipping in each $x$-axis bin (bin size: 0.1). Each error bar denotes the standard deviation of the mean values from 1000 bootstrap resamplings. Each line is a line fitted to the squares of the same color.  
\label{fig:ro}}
\end{figure*} 

\begin{figure}
\includegraphics[width=\linewidth]{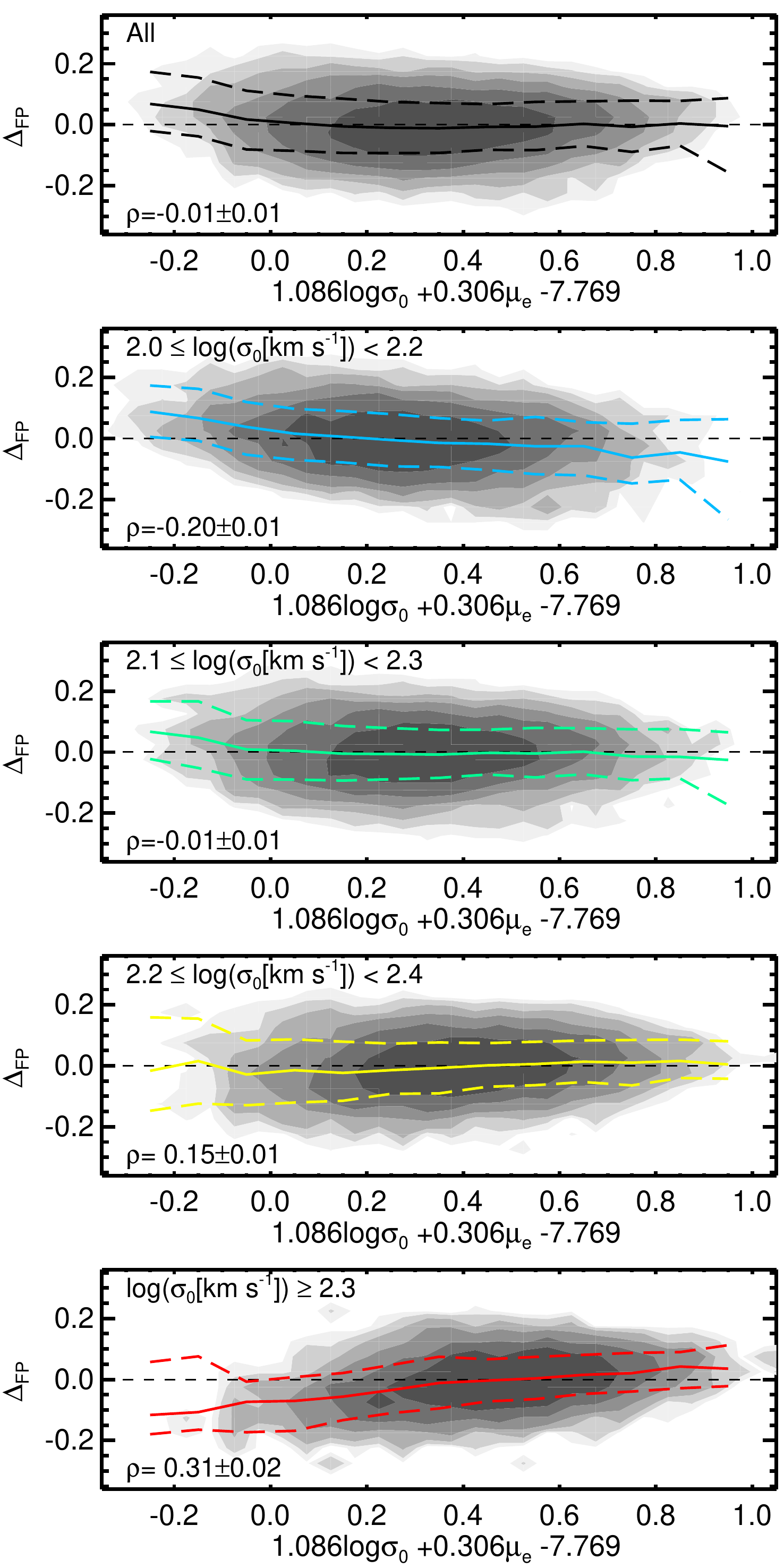}
\centering
\caption{Residuals ($\Delta_\mathrm{FP}$; Equation \ref{eq:resi}) for the $z$-band FP. Residuals for ETGs in different $\log\sigma_0$ bins are also shown in the different panels. The thin black dashed lines ($\Delta_\mathrm{FP}=0$) indicate the FP derived from the whole ETGs (edge-on view). The solid lines are the spline-fit lines that represent nonlinear continuous relations for $\Delta_\mathrm{FP}$. The thick dashed lines above and under the spline-fit lines indicate 68th percentile ranges for $\Delta_\mathrm{FP}$. In the lower left corner of each panel, the Spearman correlation coefficient is presented with its error. The levels of the contours in each panel are scaled down from the contour levels in Figure \ref{fig:FPall} by a ratio between the number of ETGs in each $\log\sigma_0$ bin and the number of the whole ETGs.
\label{fig:resi}}
\end{figure}

\begin{figure*}
\includegraphics[width=\linewidth]{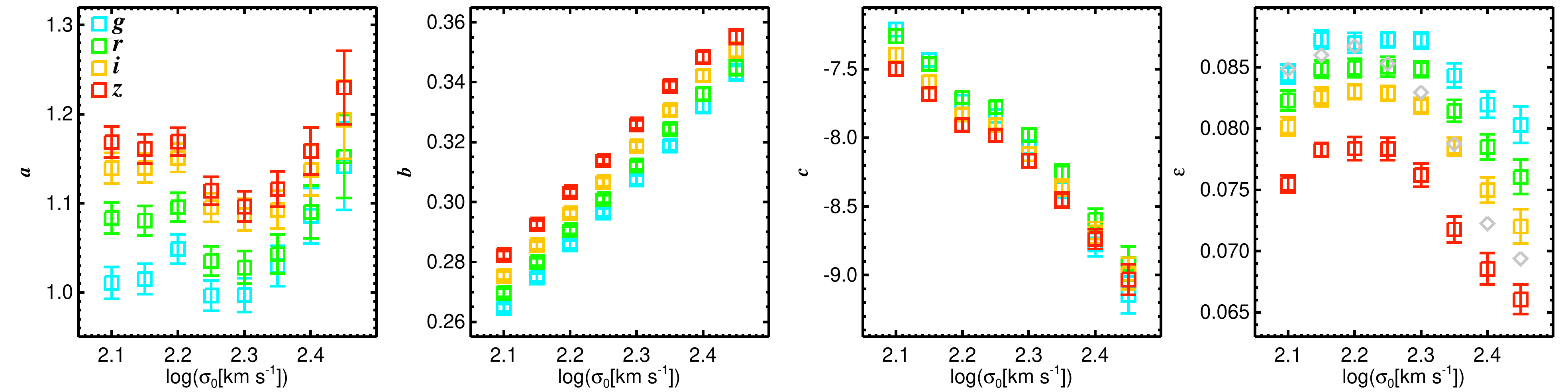}  
\centering
\caption{Coefficients of the FPs ($a$, $b$, and $c$) and intrinsic scatters ($\varepsilon$) around the FPs for ETGs in different $\log\sigma_0$ bins (see Equations \ref{eq:fp} and \ref{eq:chi}). The results for different bands are displayed in different colors. The bin size is 0.2 dex centered on each square, so that $\log\sigma_0$ bins overlap each other. The last bin contains all ETGs with $\log\sigma_0\ge2.35$. The gray diamonds in the last panel denote observed scatters around the $z$-band FPs where the observational uncertainties are not taken into account.
\label{fig:coeff}}
\end{figure*}

\section{Binning ETG Sample}\label{sec:binning}

It is possible to investigate the curved nature of the FP by dividing ETGs into several subsamples according to $\log\sigma_0$ or $\mu_e$ and inspecting the change in the coefficients of the FP fitted to each subsample. Since a consistent conclusion can be reached based on either $\log\sigma_0$ bins or $\mu_e$ bins as shown in Section \ref{sec:results}, we elected to divide ETGs mainly by $\log\sigma_0$ in order to avoid redundant processes and descriptions. We prefer the use of $\log\sigma_0$ bins for two reasons. One is that unlike $\mu_e$, $\sigma_0$ values are independent of the used bands, which means that $\sigma_0$ is a more universally applicable parameter. The other is that cutting the ETG sample with $\log\sigma_0$ leads to slightly (at most $\sim0.01$ dex) smaller scatters around the FPs than binning with $\mu_e$. Moreover, it makes the cut edge of the ETG distribution closer to perpendicular to the edge-on view direction of the FP of all ETGs. Thus, the change in the shape of the FP is displayed a little bit more prominently when ETGs are divided by $\log\sigma_0$, especially in the edge-on view of the FP.

Splitting ETGs into too small $\log\sigma_0$ bins ($<0.05$ dex) can lead to spurious plane fitting, since it forces the orthogonal direction to the fitted plane to be aligned with the $\log\sigma_0$ axis, due to the excessively narrow ETG distribution in the direction of the $\log\sigma_0$ axis. Therefore, we set the sizes of $\log\sigma_0$ bins to be larger than or equal to 0.2 dex. We note that the plane-fitting method used here, which minimizes residuals in the direction of the $\log R_e$ axis (direct fit; see Equation \ref{eq:chi}), is less affected by the $\sigma_0$ cut than the method that minimizes residuals in the orthogonal direction to the fitted plane (orthogonal fit), as shown in previous studies \citep{Bernardi2003b,Hyde2009b} and in Appendix \ref{Appendixcomp} where we describe comparisons between the different fitting methods.

We tested how much the coefficients of the FP are affected by the geometry of the ETG distribution that comes from binning the galaxies with $\log\sigma_0$. We generated 20,000 mock galaxies whose $R_e$, $\sigma_0$, and $\mu_e$ follow the $z$-band FP of all ETGs with the intrinsic scatter around the plane.\footnote{See Figure \ref{fig:FPall} and Table \ref{tb:fp} in Appendix \ref{AppendixTable}.} Then, we varied the size of the $\log\sigma_0$ bin from 0.01 to 0.6 dex and fitted planes to distributions of mock galaxies within the bins as was done for the observational data. The test was conducted for 50 sets of mock galaxies whose parameters were randomly generated in each set. Specifically, the 20,000 mock galaxies in each set were generated as follows: (1) producing mock galaxies whose $\log\sigma_0$ values exactly follow the $\log\sigma_0$ distribution for observational data shown in Figure \ref{fig:sigdist}; (2) assigning $\mu_e$ values to the mock galaxies in each $\log\sigma_0$ bin with a size of 0.05 dex, in such a way that the distribution of the assigned $\mu_e$ values follows that of observational ETGs within the $\log\sigma_0$ bin; (3) assigning $\log R_e$ values to mock galaxies, in which $\log R_e$ values follow the FP relation of the whole ETGs and have a scatter according to the scatter of the FP relation. The three parameters of the mock galaxies generated in this way have almost the same distributions (and correlations) as the observational data shown in Figure \ref{fig:sigdist}, even when the galaxies are divided into $\log\sigma_0$ bins.

The test shows that the bias in the coefficient $a$ ($\delta a/a$) of the FP due to the ETG sample cut with $\log\sigma_0$ can be minimized to less than $\sim4\%$ of the true value by applying the minimum bin size of 0.2 dex and the direct-fit method.\footnote{The bias in the coefficient $a$ is more than $\sim10\%$ of the true value, when the size of the $\log\sigma_0$ bin is less than 0.05 dex.} By contrast, the coefficient $b$ is not affected by the $\sigma_0$ cut (see also the test for the $\sigma_0$ cut in \citealt{Hyde2009b}). 

The same test was conducted for the case when the sample is divided by $\mu_e$, varying the size of the $\mu_e$ bin from 0.05 to 3.5 mag arcsec$^{-2}$. The test shows that the bias in the coefficient $b$ can be minimized to within $\sim3\%$ of the true value by adopting the minimum bin size of 1 mag arcsec$^{-2}$, while the coefficient $a$ is hardly influenced by the $\mu_e$ cut.

Since the whole range of $\log\sigma_0$ occupied by our ETGs is $\sim0.6$ dex, we overlapped $\log\sigma_0$ bins when gradual changes as a function of $\log\sigma_0$ were examined. 

Even in the case when we only use ETGs with $\log\sigma_0\gtrsim2.1$ that have little to do with the magnitude cut of $M_r\le-19.5$,\footnote{ETGs with $\log\sigma_0\sim2.1$ have the median $M_r$ of $-20.4$, and $95\%$ of them are at $M_r<-19.6$, when the magnitude cut of $M_r\le-19.5$ is not applied to the sample.} the trends of the results shown in Sections \ref{sec:results} and \ref{sec:imp} are still evident and consistent with those of the full sample with $\log\sigma_0\ge2.0$. With the fact that $\log\sigma_0=2.0$ corresponds to $M_r\sim-19.9$ for ETGs used here, it means that the magnitude cut of $M_r\le-19.5$, which was applied to our sample to make the volume-limited sample in $0.025\le z_\mathrm{spec}<0.055$, does not alter our main conclusions.
\\

\begin{figure*}
\includegraphics[width=\linewidth]{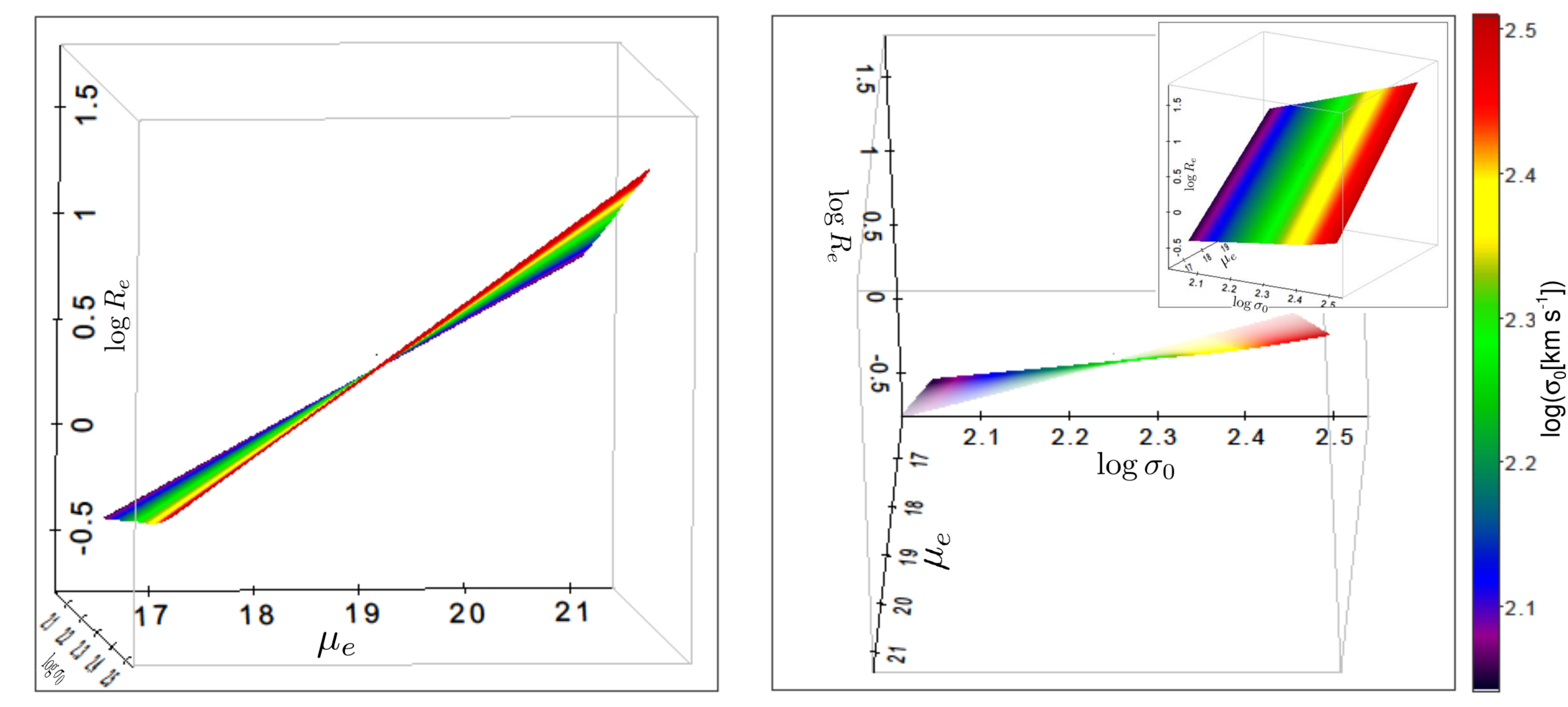} 
\centering                      
\caption{The reproduced FP in the three-dimensional parameter space, which is based on the $z$-band fitting results for different $\log\sigma_0$ bins in Figure \ref{fig:coeff}. The FP is reproduced within $2.05<\log\sigma_0<2.50$ and $16.5<\mu_e<21.0$. For the purpose of emphasizing the curvature (or variations of the coefficients), we ignored the intrinsic scatter of ETGs around the FP and only the infinitely thin surface is shown here. The surface is color-coded according to $\log\sigma_0$ values (see the color bar and the inset in the right panel). The figure shows the reproduced FP surface seen from the directions where the twist of the surface is clearly visible. 
\label{fig:3D}}
\end{figure*}

\begin{figure*}
\includegraphics[width=\linewidth]{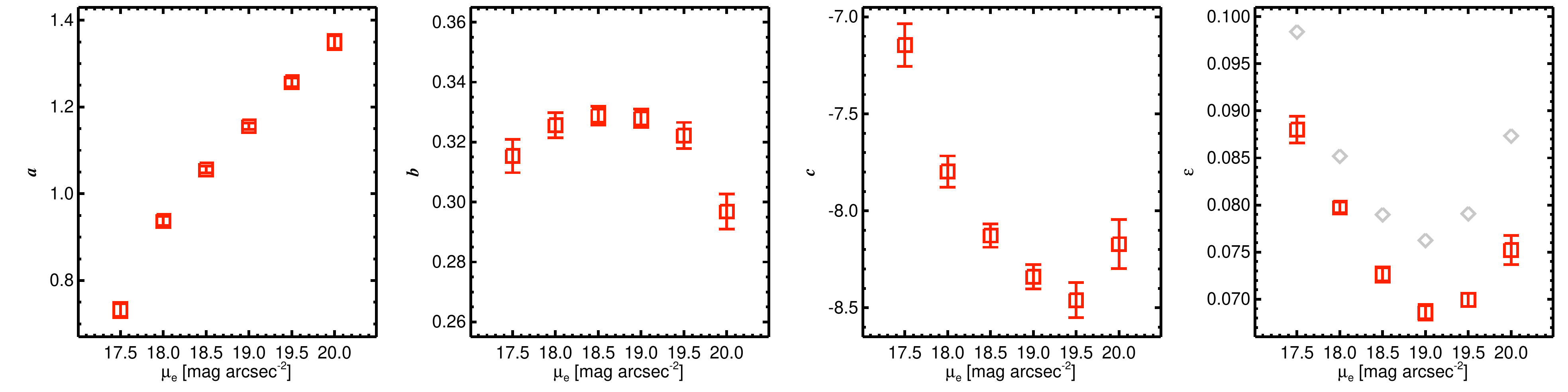}  
\centering
\caption{Coefficients of the $z$-band FPs ($a$, $b$, and $c$) and intrinsic scatters ($\varepsilon$) around the FPs for ETGs in different $\mu_e$ bins. The bin size is 1 mag arcsec$^{-2}$ centered on each square, so that $\mu_e$ bins overlap each other. The first bin and the last bin contain all ETGs with $\mu_e < 18.0$ and $\mu_e \ge 19.5$, respectively.  The gray diamonds in the last panel denote observed scatters around the FPs where the observational uncertainties are not taken into account.
\label{fig:coeffmu}}
\end{figure*}

\begin{figure}
\includegraphics[width=\linewidth]{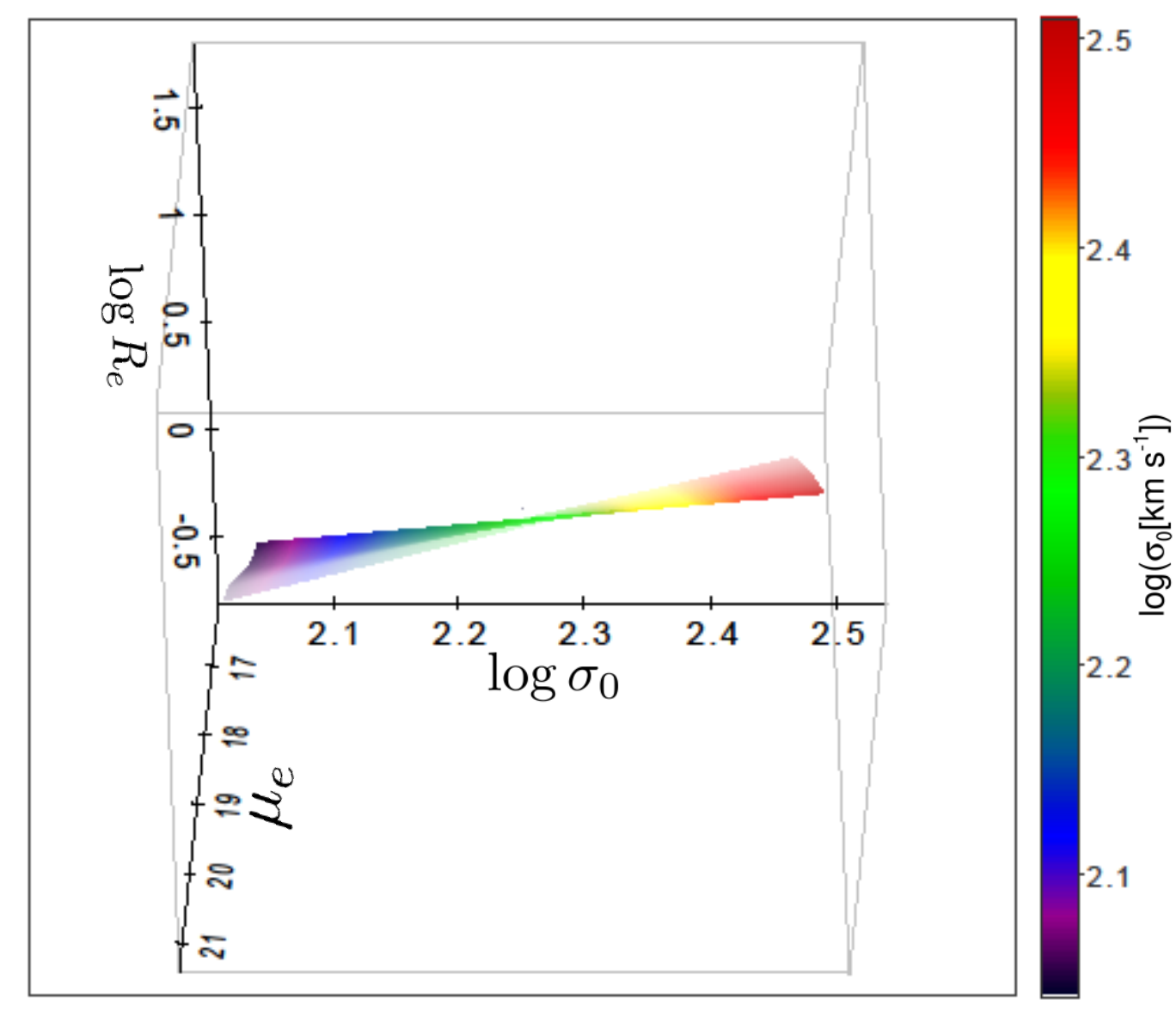} 
\centering                      
\caption{The reproduced FP in the three-dimensional parameter space, which is based on the $z$-band fitting results for different $\mu_e$ bins in Figure \ref{fig:coeffmu}. The FP is reproduced within $2.05<\log\sigma_0<2.50$ and $17.0<\mu_e<20.5$ (the range for $\mu_e$ is slightly smaller than that of Figure \ref{fig:3D}). Other descriptions about this figure are identical to those in Figure \ref{fig:3D}.
\label{fig:3Dmu}}
\end{figure}

\section{Results}\label{sec:results}

The results of this study are essentially identical regardless of the used bands. Thus, in this study we mainly show results based on the $z$ band. The coefficients of the FP and intrinsic scatter of galaxies around the FP for different $\log\sigma_0$ bins and bands are summarized in Table \ref{tb:fp} in Appendix \ref{AppendixTable}.

Figure \ref{fig:sigbin} shows ETG distributions in the three-dimensional parameter space of $\log R_e$, $\log\sigma_0$, and $\mu_e$ for three different $\log\sigma_0$ bins. For direct comparison, superposed in Figure \ref{fig:comp} are equal density contours for distributions of ETGs in the three $\log\sigma_0$ bins. In Figures \ref{fig:sigbin} and \ref{fig:comp}, all the ETG distributions are projected toward the edge-on view direction of the $z$-band FP of all ETGs. The two figures indicate that the slope and the zero-point of the FP change in the sense that the FP of ETGs with higher $\sigma_0$ has a positively steeper slope and a lower zero-point, even though the FP of all ETGs is close to a plane. This trend is also evident in Figure \ref{fig:ro} which directly shows how the slope and the zero-point of the projected ETG distribution change as a function of $\sigma_0$.  

We analyzed residuals of the FP and display the result in Figure \ref{fig:resi}, which shows residuals ($\Delta_\mathrm{FP}$) for the $z$-band FP. The residual $\Delta_\mathrm{FP}$ is defined as
\begin{equation}
\Delta_\mathrm{FP}=\log R_e - a\log\sigma_0 - b\mu_e - c.
\label{eq:resi}
\end{equation}
In this analysis for the $z$-band case, we used the fixed values of $a=1.086$, $b=0.306$, and $c=-7.769$ (the coefficients correspond to the $z$-band FP for all ETGs). Also shown in the figure are the residuals for ETGs in different $\log\sigma_0$ bins. We calculated the Spearman correlation coefficients (hereafter $\rho$) to examine deviations of the residuals from $\Delta_\mathrm{FP}=0$ quantitatively. They are shown in the lower left corners of the panels in Figure \ref{fig:resi} with their errors that were computed from 1000 bootstrap resamplings.

In the case when using the whole ETGs shown in the top panel of Figure \ref{fig:resi}, the average $\Delta_\mathrm{FP}$ is very close to 0 across all ranges of $x$-axis (i.e., $1.086\,\log\sigma_0 + 0.306\,b\mu_e - 7.769$), except at the very low part of the $x$-axis ($\lesssim-0.1$), in which ETGs with low $\log\sigma_0$ are the dominant population. Moreover, $\rho$ is nearly zero ($\rho=-0.01\pm0.01$). The residuals and $\rho$ indicate that the FP was properly fitted to the ETG distribution in the three-dimensional parameter space.

Being consistent with the previous figures of this paper, Figure \ref{fig:resi} shows that residuals for ETGs in the low or high $\log\sigma_0$ bins deviate from $\Delta_\mathrm{FP}=0$ in such a way that ETGs in higher (lower) $\log\sigma_0$ bins have higher (lower) $\rho$. For instance, ETGs with $2.0\le\log\sigma_0<2.2$ have $\rho=-0.20\pm0.01$, whereas those with $\log\sigma_0\ge2.3$ have $\rho=0.31\pm0.02$. Thus, the deviation trends for residuals in the two extreme bins are significantly opposite to each other, which implies that the exact shape of the FP is close to a twisted surface.

In Figure \ref{fig:coeff}, we display coefficients of the FPs ($a$, $b$, and $c$ in Equation \ref{eq:fp}) and intrinsic scatters around the FPs for ETGs in different $\log\sigma_0$ bins. The coefficient of $\log\sigma_0$ ($a$ coefficient) does not show a significant change as a function of $\log\sigma_0$. Although the highest $\sigma_0$ bin shows a slightly higher $a$ coefficient than the other bins, the differences in $a$ coefficient between the $\log\sigma_0$ bins are roughly within two to three times the error values. By contrast, the coefficient of $\mu_e$ ($b$ coefficient) shows a significant trend that ETGs with higher $\sigma_0$ have higher $b$ coefficients, so that in the case of the $z$ band $b$ is $0.282\pm0.001$ for ETGs with $2.0\leq\log\sigma_0<2.2$, while it is $0.355\pm0.002$ for ETGs with $\log\sigma_0\geq2.35$. Therefore, the slope change in the FP as a function of $\sigma_0$ shown in Figures \ref{fig:sigbin}, \ref{fig:comp}, and \ref{fig:ro} is mainly due to the variation of the $b$ coefficient of the FP over different $\log\sigma_0$ bins.

The coefficient $c$ also changes evidently in the sense that ETGs with higher $\sigma_0$ have a lower $c$, which is consistent with the zero-point variation in the FP shown in Figures \ref{fig:sigbin}, \ref{fig:comp}, and \ref{fig:ro}. For example, in the case of the $z$ band, $c$ is $-7.500\pm0.049$ for ETGs with $2.0\leq\log\sigma_0<2.2$, whereas it is $-9.033\pm0.111$ for ETGs with $\log\sigma_0\geq2.35$. 

The last panel of Figure \ref{fig:coeff} shows that ETGs with higher $\sigma_0$ have smaller intrinsic scatters around the FPs at $\log\sigma_0\gtrsim2.2$, so that in the case of the $z$ band $\varepsilon$ is $0.066$ for ETGs with $\log\sigma_0\geq2.35$, whereas it is $0.078$ at $\log\sigma_0\sim2.2$. We note that the trend found in $\varepsilon$ is also detected in the observed scatters around the FPs where the observational uncertainties are not considered (the gray diamonds in the last panel of Figure \ref{fig:coeff}).

The FP depends on the band. The coefficients $a$ and $b$ are higher and $c$ is lower at the redder bands. In addition, the intrinsic scatter is smaller at the redder bands \citep{YP2020}.

We reproduced the FP based on the $z$-band fitting results in Figure \ref{fig:coeff}, which is displayed in the three-dimensional parameter space shown in Figure \ref{fig:3D}. We interpolated the coefficients of the FP to produce a smooth and continuous surface. In Figure \ref{fig:3D}, we ignored the intrinsic scatter of ETGs around the FP, in order to highlight the curvature of the surface (or variations of the coefficients). The figure shows that the FP is roughly flat, but in the strict sense, it is a twisted surface.

Although we investigated the curved nature of the FP by dividing ETGs into $\log\sigma_0$ bins, Figure \ref{fig:3D} demonstrates that the changes in the coefficients of the FP can also be detected when ETGs are separated into different $\mu_e$ bins. Figure \ref{fig:coeffmu} shows coefficients of the $z$-band FPs and intrinsic scatters around the FPs for ETGs in different $\mu_e$ bins, which are summarized in Table \ref{tb:fpmu} in Appendix \ref{AppendixTable} with the results for the other bands. In this case, the $a$ coefficient of the FP shows a significant trend that ETGs with higher $\mu_e$ (fainter surface brightness) have higher $a$ coefficients, while the $b$ coefficient of the FP does not show a significant variation as a function of $\mu_e$ compared to the trend in the $a$ coefficient, as naturally expected from the twisted surface shown in Figure \ref{fig:3D}. 

In Figure \ref{fig:3Dmu}, we reproduced the FP based on the results in Figure \ref{fig:coeffmu} in the same way as the construction method for the FP in Figure \ref{fig:3D}. The twisted shape of the reproduced FP in Figure \ref{fig:3Dmu} is almost identical to that in the right panel of Figure \ref{fig:3D}, which indicates that the consistent picture on the warped nature of the FP is also obtained using the variations of the FP coefficients for ETGs in different $\mu_e$ bins.
\\

\begin{figure}
\includegraphics[width=\linewidth]{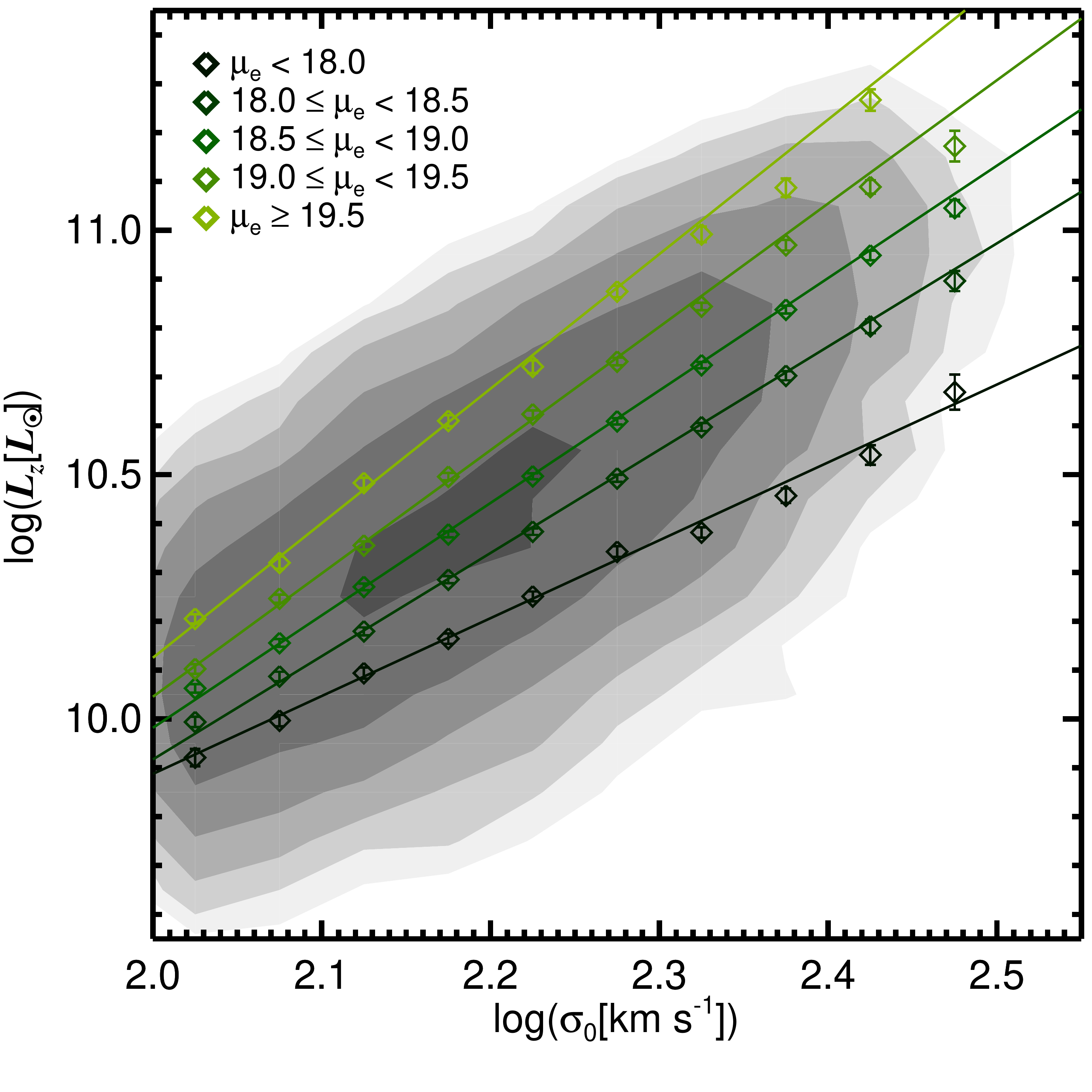}  
\centering
\caption{$L$--$\sigma_0$ relations (Faber--Jackson relations) for ETGs with different $\mu_e$ (the color lines). The parameters used here are based on the $z$ band. Each diamond represents the mean $\log L_z$ after 3$\sigma$ clipping in each $\log\sigma_0$ bin (bin size: 0.05 dex). Each error bar denotes the standard deviation of the mean values from 1000 bootstrap resamplings. The levels of the contours indicate 12, 24, 48, 96, 192, and 384 galaxies in 2D bins whose sizes in the $x$- and $y$-axes are 0.05 and 0.1, respectively. We note that ETGs with $M_r>-19.5$ are included in this figure, as well as in the fitting of the $L$--$\sigma_0$ relations. 
\label{fig:lvd}}
\end{figure} 

\begin{figure}
\includegraphics[width=\linewidth]{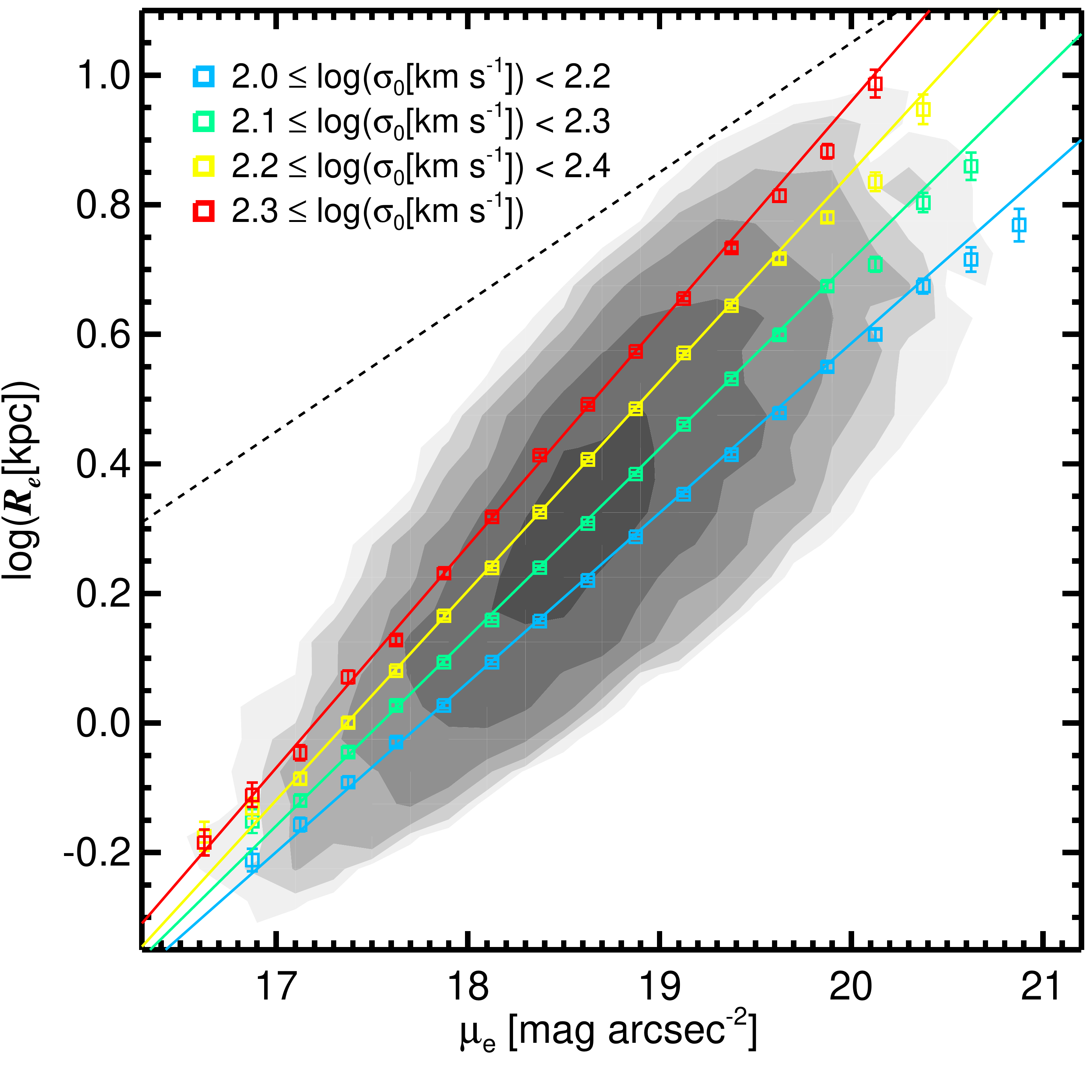}  
\centering
\caption{$R_e$--$\mu_e$ relations (Kormendy relations) for ETGs with different $\log\sigma_0$ (the color lines). The parameters used here are based on the $z$ band. Each square denotes the mean $\log R_e$ after 3$\sigma$ clipping in each $\mu_e$ bin (bin size: 0.25). Each error bar indicates the standard deviation of the mean values from 1000 bootstrap resamplings. The black dashed line represents the slope for galaxies that have the same luminosity. The levels of the contours indicate 6, 12, 24, 48, 96, and 192 galaxies in 2D bins whose sizes in the $x$- and $y$-axes are 0.2 and 0.05, respectively.
\label{fig:rmu}}
\end{figure} 

\begin{figure}
\includegraphics[width=\linewidth]{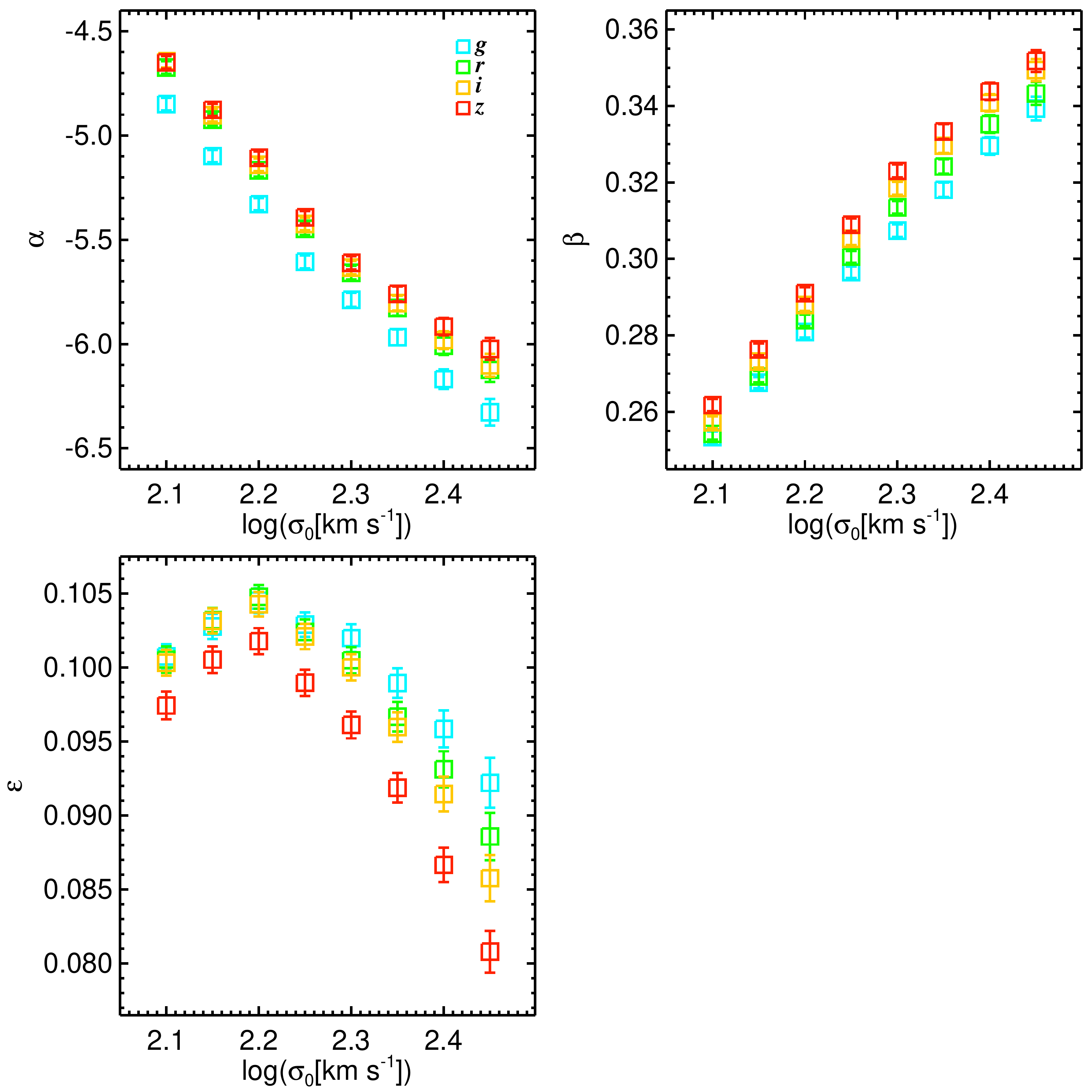}  
\centering
\caption{Coefficients ($\alpha$ and $\beta$) of the $R_e$--$\mu_e$ relations and intrinsic scatters ($\varepsilon$) around the relations for ETGs in different $\log\sigma_0$ bins (see Equations \ref{eq:chi2} and \ref{eq:rm}). The results for different bands are displayed in different colors. The bin size is 0.2 dex centered on each square, so that $\log\sigma_0$ bins overlap each other. The last bin contains all ETGs with $\log\sigma_0\ge2.35$. 
\label{fig:coeff_rmu}}
\end{figure}

\section{Implication for Formation of ETG\lowercase{s}}\label{sec:imp}
In this section, we discuss the origin of the warped nature of the FP and its implication for the formation of ETGs. We begin the discussion with the $L$--$\sigma_0$ relation (Faber--Jackson relation; \citealt{Faber1976}), which is described by 
\begin{equation}
\log L=p+q\log\sigma_0,
\label{eq:ls}
\end{equation}
where $L$ is the luminosity of ETGs in each band. The units of $L$ used in this study are the solar luminosities ($L_\odot$) in the four SDSS bands \citep{Wilmer2018}. We included 496 ETGs with $M_r>-19.5$ to fit the $L$--$\sigma_0$ relation. Thus, the total number of ETGs used to fit the relation is 16,779. The coefficients of the $L$--$\sigma_0$ relation and the intrinsic scatter around the relation for different $\mu_e$ bins and bands are summarized in Table \ref{tb:ls} in Appendix \ref{AppendixTable}.

Figure \ref{fig:lvd} displays $z$-band $L$--$\sigma_0$ relations for ETGs with different $\mu_e$. The figure shows that ETGs with lower surface brightness (higher values of $\mu_e$) have higher $\log L_z$ (or higher stellar masses) for a given $\log\sigma_0$. In addition, the slope of the $L$--$\sigma_0$ relation ($q$ in Equation \ref{eq:ls}) depends on $\mu_e$ in the sense that ETGs with lower surface brightness have higher slopes. For example, ETGs with $\mu_e\ge19.5$ have $q=2.75\pm0.04$, while ETGs with $\mu_e<18.0$ have $q=1.59\pm0.04$.\footnote{The dependence of the slope on $\mu_e$ is consistent with (but not exactly the same as) the result in \citet{Kormendy2013} in the sense that core ellipticals, whose central surface brightness profiles are shallow (hence lower surface brightness), have a higher slope in the $L$--$\sigma_0$ relation, whereas coreless ellipticals, which show steep surface brightness profiles in galaxy centers (hence higher surface brightness), have a lower slope in the $L$--$\sigma_0$ relation.}

It is known that dry mergers do not substantially increase velocity dispersions of galaxies in contrast to stellar masses, which are directly proportional to the amount of masses accreted through dry mergers \citep{Nipoti2003,Robertson2006,Lauer2007, Hopkins2009,Hilz2012,Kormendy2013}. Moreover, dry mergers, especially minor mergers, are able to effectively increase sizes of ETGs more than stellar masses \citep{Lauer2007,Lopez2012,Oogi2013,Oogi2016,Yoon2017}, which causes low surface brightness in merger remnants. Therefore, the two effects of dry mergers can explain the trend in the $L$--$\sigma_0$ relation that ETGs with lower surface brightness have higher luminosities for a given $\log\sigma_0$.

More massive/luminous ETGs are expected to have experienced more (dry) mergers in their formation histories \citep{Desroches2007,Bernardi2011a,Bernardi2011b,Yoon2017,Yoon2022,YL2020,OLeary2021}. Thus, the dry merger effects imprinted in galaxy properties should be prominent in ETGs with higher luminosities \citep{Desroches2007}. This mass/luminosity dependence is manifested in the key feature of the $L$--$\sigma_0$ relation that the differences in $\log L$ between different $\mu_e$ bins for a given $\log\sigma_0$ increase gradually as $\log L$ (or $\log\sigma_0$) rises (in other words, the luminosity variation as a function of $\mu_e$ is more substantial in ETGs with higher $\log\sigma_0$).

 According to the definition of $\mu_e$ (Equation \ref{eq:mu}), this key feature in the $L$--$\sigma_0$ relation is equivalent to the fact that the slope of the $R_e$--$\mu_e$ relation is higher in ETGs with higher $\log\sigma_0$. Figure \ref{fig:rmu} displays $z$-band $R_e$--$\mu_e$ relations (Kormendy relation; \citealt{Kormendy1977}) for ETGs with different $\log\sigma_0$. The $R_e$--$\mu_e$ relation is described by 
\begin{equation}
\log R_e=\alpha+\beta\mu_e.
\label{eq:rm}
\end{equation}
The coefficients of the $R_e$--$\mu_e$ relation and the intrinsic scatter around the relation for different $\log\sigma_0$ bins and bands are summarized in Table \ref{tb:rm} in Appendix \ref{AppendixTable} as well as in Figure \ref{fig:coeff_rmu}.  As mentioned above, ETGs with higher $\sigma_0$ have higher slopes ($\beta$) and lower zero-points. For example, in the case of the $z$ band, $\beta$ is $0.262\pm0.002$ for ETGs with $2.0\leq\log\sigma_0<2.2$, while it is $0.352\pm0.003$ for ETGs with $\log\sigma_0\geq2.35$. Moreover, the intrinsic scatter around the $R_e$--$\mu_e$ relation shows a trend that it is smaller at higher $\sigma_0$ at $\log\sigma_0\gtrsim2.2$.

Since the $R_e$--$\mu_e$ relation is a projection of the FP toward the direction of the $\log\sigma_0$ axis, the $\sigma_0$ dependence in the coefficients of the $R_e$--$\mu_e$ relation and the intrinsic scatter around the relation is directly reflected in the different FPs at different $\log\sigma_0$ bins described in Section \ref{sec:results}. Similarly, the key feature in the $L$--$\sigma_0$ relation also indicates that the slope of the $R_e$--$\log\sigma_0$ relation is positively steeper in ETGs with lower surface brightness (higher $\mu_e$) as can be inferred from Figure \ref{fig:rmu}, and this $\mu_e$ dependence in the $R_e$--$\log\sigma_0$ relation is reflected in the shapes of the FP. Therefore, the warped nature of the FP of ETGs can be explained by dry merger effects that are imprinted more prominently in properties of ETGs with higher luminosities.
\\

\section{Comparison with Other Studies}\label{sec:comp}

\citet{D'Onofrio2008} and \citet{Nigoche-Netro2009} showed that the coefficients of the FP depend on the luminosity/magnitude cut applied to the sample, and they found that this dependence is from the geometry of the ETG distribution that the magnitude cut brings about. In particular, selecting ETGs within a narrow range of the luminosity/magnitude is able to cause the coefficients of the FP to be distorted substantially, since the narrow cut in the magnitude forces ETGs to have a fixed slope in the $R_e$--$\mu_e$ relation (by the definition of $\mu_e$), as indicated by the dashed line in Figure \ref{fig:rmu}. 

The study for the FP in \citet{Hyde2009b} suggested possible curvature in the FP at the low-mass/small-size end of ETGs, which is contrary to our results that evidently show continuous changes in the coefficients of the FP over the full range of $\sigma_0$.

Using the orthogonal fit to derive the FP, \citet{Bernardi2003b} and \citet{Gargiulo2009} found that the slope of the FP changes when $\sigma_0$ cuts are applied to the sample. They argued that this slope change is merely an artifact from a selection bias in the geometry of the ETG distribution caused by the $\sigma_0$ cut. The $\sigma_0$ cut applied to ETG samples can induce a bias in the coefficient of $\log\sigma_0$ (see Appendix \ref{Appendixcomp} and the test in \citealt{Hyde2009b}). However, the coefficient of $\mu_e$, which is found to change when ETGs are divided by $\log\sigma_0$ here, is immune to binning ETGs with $\sigma_0$. We also used the adequate minimum bin size and the direct-fit method that can minimize the bias in the derived coefficients to within $\sim4\%$ of the true values (Section \ref{sec:binning}). Furthermore, the same curved nature of the FP can also be detected by cutting samples with $\mu_e$. Therefore, our discovery that the FP of ETGs is a curved surface is not a simple artifact from the sample cut applied to the ETG sample.

\citet{Samir2020} found that BCGs (higher $\sigma_0$) and isolated ellipticals (lower $\sigma_0$) follow different FPs, mainly due to the significant difference in the coefficient of $\mu_e$ (and the slope of the $R_e$--$\mu_e$ relation). They also showed that BCGs have a smaller scatter around the FP than isolated ellipticals. Therefore, our results are consistent with those of \citet{Samir2020}.

Several studies investigated the intrinsic scatter around the FP for different ETG populations. For example, \citet{YP2020} showed that ETGs with old stellar populations have a small intrinsic scatter around the FP compared with young ETGs. \citet{Bernardi2020} found that slow-rotating ellipticals have a smaller intrinsic scatter around the FP than fast-rotating ellipticals and lenticular galaxies. The main results of these two studies are consistent with ours, since ETGs with high $\sigma_0$, which are found to show a small intrinsic scatter around the FP, have relatively older stellar populations \citep{Graves2009a,Graves2009b,YP2020} and rotate more slowly \citep{Graham2018,Yoon2022} than ETGs with low $\sigma_0$.
\\

\section{Summary}\label{sec:summary}

We investigated the warped nature of the FP of ETGs by inspecting the dependence of the coefficients of the FP mainly on $\sigma_0$ and additionally on $\mu_e$. For this, we used $16,283$ ETGs with $M_r\le-19.5$ in $0.025\le z_\mathrm{spec}<0.055$ from SDSS data. By doing so, we found that the FP of ETGs is not a plane in the strict sense but is a curved surface that varies its orthogonal direction to the surface as $\sigma_0$ or $\mu_e$ changes. 

When ETGs are separated into subsamples with different $\log\sigma_0$, the coefficient of $\mu_e$ of the FP rises while the zero-point of the FP falls at higher $\sigma_0$. In the case of the $z$ band, the coefficient of $\mu_e$ and the zero-point of the FP are $0.282\pm0.001$ and $-7.500\pm0.049$, respectively, for ETGs with $2.0\leq\log\sigma_0<2.2$, whereas they are $0.355\pm0.002$ and $-9.033\pm0.111$, respectively, for ETGs with $\log\sigma_0\geq2.35$. The consistent picture on the curved nature of the FP is also reached by examining variations of the FP coefficients as a function of $\mu_e$. By investigating scaling relations that are projections of the FP, we suggest that the warped nature of the FP may be due to dry merger effects that are imprinted more in ETGs with higher stellar masses (or higher luminosities). 
\\

\begin{acknowledgments}
This work was supported by a KIAS Individual Grant PG016904 at the Korea Institute for Advanced Study. This research was supported by the Korea Astronomy and Space Science Institute under the R\&D program (Project No. 2022-1-830-05), supervised by the Ministry of Science and ICT.
\end{acknowledgments}

\appendix

\section{Coefficients of the FP\lowercase{s} from Other FP Fitting Methods}\label{Appendixcomp}

\begin{figure*}
\includegraphics[scale=0.50]{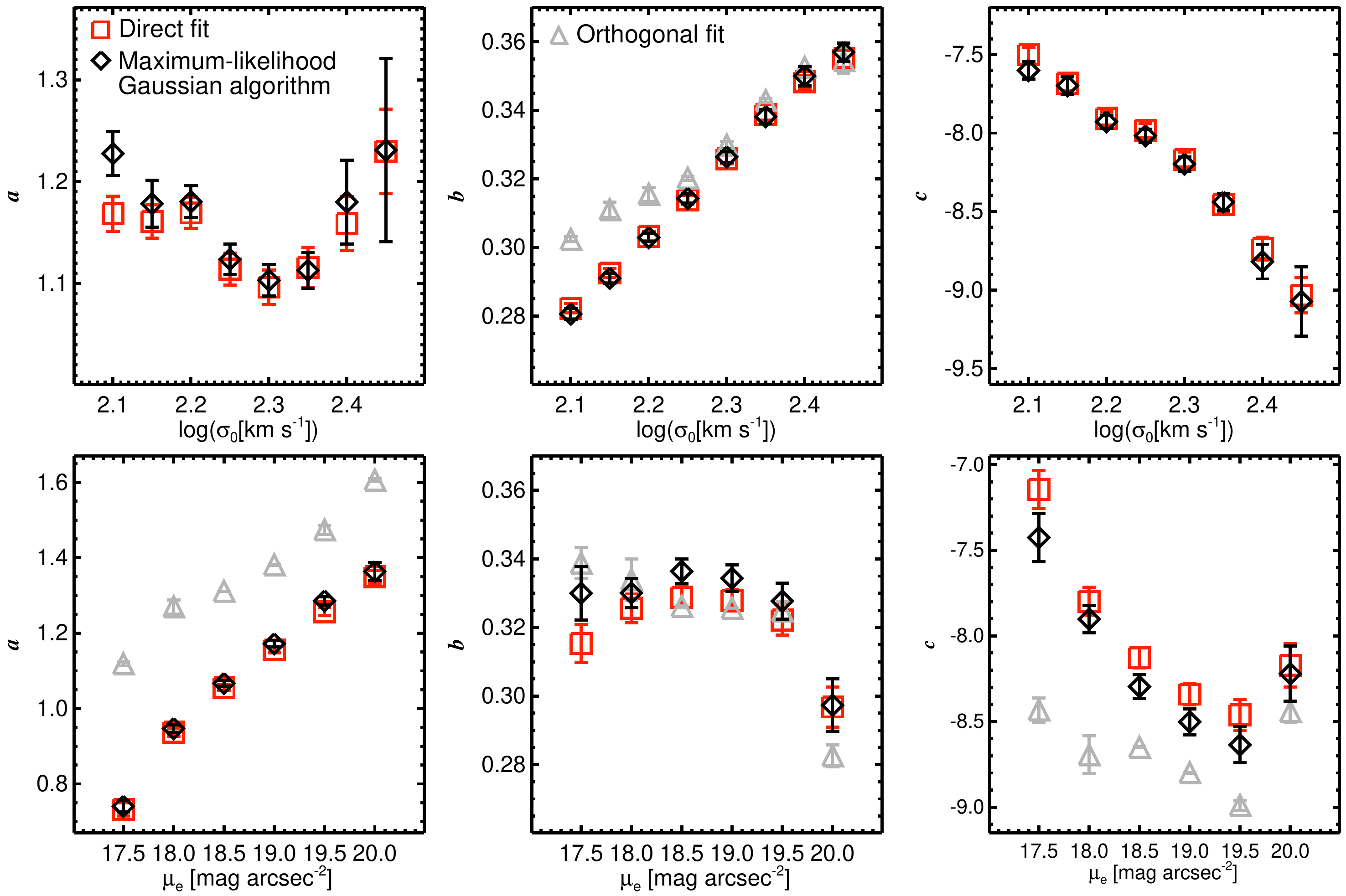}  
\centering
\caption{Coefficients of the $z$-band FPs for ETGs in different $\log\sigma_0$ bins (top panels) and $\mu_e$ bins (bottom panels). The coefficients were derived from three different fitting methods: (1) direct-fit method in Section \ref{sec:fit}, (2) maximum likelihood Gaussian algorithm to minimize $\Delta_\mathrm{FP}^2$ \citep{Saglia2001,Bernardi2003b}, and (3) orthogonal-fit method minimizing $\Delta_o$ \citep{deGraaff2020}. The coefficients $a$ and $c$ derived from the orthogonal-fit method are highly biased when ETGs are divided by $\log\sigma_0$, so that they are not displayed in the figure. The maximum likelihood Gaussian method is greatly consistent with the direct-fit method in Section \ref{sec:fit} in terms of both coefficient values and trends.
\label{fig:coeff_diff}}
\end{figure*} 

In this appendix, we test the direct-fit method presented in Section \ref{sec:fit} by comparing it with other FP fitting methods. Firstly, we derived coefficients of the FP by minimizing the mean absolute orthogonal deviations from the plane \citep{deGraaff2020,deGraaff2021},
\begin{equation}
\Delta_o=\frac{|\Delta_\mathrm{FP}|}{\sqrt{1+a^2+b^2}},
\label{eq:ortho}
\end{equation}
where $\Delta_\mathrm{FP}$ is defined in Equation \ref{eq:resi}. This is an orthogonal-fit method. We note that results from minimizing $\Delta_o^2$ are almost identical to those from minimizing $\Delta_o$. 

The coefficients of the $z$-band FP for all ETGs that are derived from the orthogonal fit are $a=1.379$, $b=0.317$, and $c=-8.632$. The coefficient $a$ is $\sim27\%$ larger than that from the direct fit, whereas $b$ is similar, as reported in \citet{Bernardi2003b}. When the ETG sample is divided into several $\log\sigma_0$ bins with a size of $0.2$ dex as in the main text, the derived $a$ coefficients are $\sim2.5$. As the bin size increases, the coefficient $a$ converges to $a\sim1.4$. This means that the coefficient $a$ is largely biased owing to the $\sigma_0$ cut when the orthogonal-fit method is used, as also reported in \citet{Hyde2009b}. The coefficient $b$, however, is not seriously affected by the $\sigma_0$ cut, so that values of $b$ and their trend as a function of $\log\sigma_0$ are similar to those from the direct fit, as shown in Figure \ref{fig:coeff_diff}, which displays coefficients of the $z$-band FPs derived from different fitting methods.

By contrast, the coefficients of the FP from the orthogonal fit are not severely biased when ETGs are divided by $\mu_e$ as shown in the bottom panels of Figure \ref{fig:coeff_diff}. Thus, in those panels, the FP coefficients from the orthogonal fit follow the trends of other methods, which indicates that a similar warped nature of the FP is detected by the orthogonal-fit method in this case.

The second method we used to derive coefficients of the FP is the maximum likelihood Gaussian algorithm to minimize $\Delta_\mathrm{FP}^2$. Details about this method are described in \citet{Saglia2001} and \citet{Bernardi2003b}. We found that the coefficients of the $z$-band FP for all ETGs from this method are $a=1.099$, $b=0.304$, and $c=-7.763$, which indicates that the coefficients of the FP differ from those of the direct-fit method in Section \ref{sec:fit} only by less than $\sim1\%$. Furthermore, as shown in Figure \ref{fig:coeff_diff}, this method is essentially in agreement with the direct-fit method in terms of both coefficient values and trends, even when the sample is finely divided by $\log\sigma_0$ and $\mu_e$. Therefore, the warped nature of the FP found in this study is identically detected through the maximum likelihood Gaussian method minimizing $\Delta_\mathrm{FP}^2$.
\\

\section{Coefficients of the FP\lowercase{s}, $L$--$\sigma_0$ Relations, and $R_{\lowercase{e}}$--$\mu_{\lowercase{e}}$ Relations}\label{AppendixTable}
Here we present four tables. Table \ref{tb:fp} lists the coefficients of the FP (Equation \ref{eq:fp}) and the intrinsic scatter of ETGs around the FP  for different $\log\sigma_0$ bins and bands. Table \ref{tb:fpmu} lists the coefficients of the FP and the intrinsic scatter of ETGs around the FP for different $\mu_e$ bins and bands. Table \ref{tb:ls} lists the coefficients of the $L$--$\sigma_0$ relation (Equation \ref{eq:ls}) and the intrinsic scatter around the relation for different $\mu_e$ bins and bands. Table \ref{tb:rm} lists the coefficients of the $R_e$--$\mu_e$ relation (Equation \ref{eq:rm}) and the intrinsic scatter around the relation for different $\log\sigma_0$ bins and bands.

\begin{deluxetable*}{crccccc}
\tablecaption{Coefficients of the FP and Intrinsic Scatter around the FP for Different $\log\sigma_0$ Bins\label{tb:fp}}
\tabletypesize{\scriptsize}
\tablehead{\colhead{Band} & \colhead{Category} & \colhead{$a$} & \colhead{$b$} & \colhead{$c$} & \colhead{$\varepsilon$} & \colhead{$N$}
}
\startdata
$g$&All&$0.966\pm0.006$&$0.287\pm0.001$&$-7.578\pm0.025$&$0.087\pm0.001$&16,283\\
$g$&$2.00\leq\log\sigma_0<2.20$&$1.011\pm0.018$&$0.265\pm0.001$&$-7.212\pm0.049$&$0.084\pm0.001$& 8010\\
$g$&$2.05\leq\log\sigma_0<2.25$&$1.015\pm0.017$&$0.275\pm0.001$&$-7.434\pm0.047$&$0.087\pm0.001$& 8951\\
$g$&$2.10\leq\log\sigma_0<2.30$&$1.049\pm0.016$&$0.286\pm0.001$&$-7.737\pm0.045$&$0.087\pm0.001$& 9324\\
$g$&$2.15\leq\log\sigma_0<2.35$&$0.996\pm0.017$&$0.297\pm0.001$&$-7.841\pm0.046$&$0.087\pm0.001$& 8805\\
$g$&$2.20\leq\log\sigma_0<2.40$&$0.997\pm0.019$&$0.308\pm0.001$&$-8.068\pm0.052$&$0.087\pm0.001$& 7522\\
$g$&$2.25\leq\log\sigma_0<2.45$&$1.030\pm0.023$&$0.319\pm0.002$&$-8.376\pm0.062$&$0.084\pm0.001$& 5595\\
$g$&$2.30\leq\log\sigma_0<2.50$&$1.086\pm0.031$&$0.332\pm0.002$&$-8.778\pm0.084$&$0.082\pm0.001$& 3603\\
$g$&$\log\sigma_0\geq2.30$&$1.090\pm0.029$&$0.332\pm0.002$&$-8.795\pm0.081$&$0.082\pm0.001$& 3642\\
$g$&$\log\sigma_0\geq2.35$&$1.142\pm0.049$&$0.343\pm0.003$&$-9.144\pm0.133$&$0.080\pm0.001$& 1869\\
\hline
$r$&All&$1.014\pm0.006$&$0.293\pm0.001$&$-7.564\pm0.024$&$0.085\pm0.001$&16,283\\
$r$&$2.00\leq\log\sigma_0<2.20$&$1.083\pm0.017$&$0.270\pm0.001$&$-7.262\pm0.048$&$0.082\pm0.001$& 8010\\
$r$&$2.05\leq\log\sigma_0<2.25$&$1.080\pm0.017$&$0.280\pm0.001$&$-7.462\pm0.046$&$0.085\pm0.001$& 8951\\
$r$&$2.10\leq\log\sigma_0<2.30$&$1.096\pm0.016$&$0.291\pm0.001$&$-7.707\pm0.044$&$0.085\pm0.001$& 9324\\
$r$&$2.15\leq\log\sigma_0<2.35$&$1.035\pm0.017$&$0.301\pm0.001$&$-7.779\pm0.045$&$0.085\pm0.001$& 8805\\
$r$&$2.20\leq\log\sigma_0<2.40$&$1.028\pm0.018$&$0.312\pm0.001$&$-7.981\pm0.050$&$0.085\pm0.001$& 7522\\
$r$&$2.25\leq\log\sigma_0<2.45$&$1.043\pm0.022$&$0.324\pm0.002$&$-8.255\pm0.059$&$0.081\pm0.001$& 5595\\
$r$&$2.30\leq\log\sigma_0<2.50$&$1.090\pm0.030$&$0.336\pm0.002$&$-8.597\pm0.079$&$0.079\pm0.001$& 3603\\
$r$&$\log\sigma_0\geq2.30$&$1.095\pm0.028$&$0.336\pm0.002$&$-8.614\pm0.076$&$0.079\pm0.001$& 3642\\
$r$&$\log\sigma_0\geq2.35$&$1.152\pm0.047$&$0.345\pm0.003$&$-8.918\pm0.124$&$0.076\pm0.001$& 1869\\
\hline
$i$&All&$1.067\pm0.006$&$0.299\pm0.001$&$-7.691\pm0.024$&$0.082\pm0.001$&16,283\\
$i$&$2.00\leq\log\sigma_0<2.20$&$1.139\pm0.017$&$0.275\pm0.001$&$-7.394\pm0.048$&$0.080\pm0.001$& 8010\\
$i$&$2.05\leq\log\sigma_0<2.25$&$1.140\pm0.016$&$0.286\pm0.001$&$-7.596\pm0.046$&$0.083\pm0.001$& 8951\\
$i$&$2.10\leq\log\sigma_0<2.30$&$1.151\pm0.016$&$0.296\pm0.001$&$-7.831\pm0.044$&$0.083\pm0.001$& 9324\\
$i$&$2.15\leq\log\sigma_0<2.35$&$1.095\pm0.016$&$0.307\pm0.001$&$-7.909\pm0.044$&$0.083\pm0.001$& 8805\\
$i$&$2.20\leq\log\sigma_0<2.40$&$1.087\pm0.018$&$0.319\pm0.001$&$-8.120\pm0.049$&$0.082\pm0.001$& 7522\\
$i$&$2.25\leq\log\sigma_0<2.45$&$1.093\pm0.021$&$0.331\pm0.002$&$-8.364\pm0.058$&$0.079\pm0.001$& 5595\\
$i$&$2.30\leq\log\sigma_0<2.50$&$1.137\pm0.028$&$0.342\pm0.002$&$-8.690\pm0.076$&$0.075\pm0.001$& 3603\\
$i$&$\log\sigma_0\geq2.30$&$1.138\pm0.027$&$0.342\pm0.002$&$-8.698\pm0.073$&$0.075\pm0.001$& 3642\\
$i$&$\log\sigma_0\geq2.35$&$1.194\pm0.044$&$0.351\pm0.002$&$-8.990\pm0.117$&$0.072\pm0.001$& 1869\\
\hline
$z$&All&$1.086\pm0.006$&$0.306\pm0.001$&$-7.769\pm0.024$&$0.078\pm0.001$&16,283\\
$z$&$2.00\leq\log\sigma_0<2.20$&$1.169\pm0.017$&$0.282\pm0.001$&$-7.500\pm0.049$&$0.075\pm0.001$& 8010\\
$z$&$2.05\leq\log\sigma_0<2.25$&$1.161\pm0.016$&$0.293\pm0.001$&$-7.683\pm0.046$&$0.078\pm0.001$& 8951\\
$z$&$2.10\leq\log\sigma_0<2.30$&$1.169\pm0.016$&$0.303\pm0.001$&$-7.906\pm0.044$&$0.078\pm0.001$& 9324\\
$z$&$2.15\leq\log\sigma_0<2.35$&$1.114\pm0.016$&$0.314\pm0.001$&$-7.983\pm0.044$&$0.078\pm0.001$& 8805\\
$z$&$2.20\leq\log\sigma_0<2.40$&$1.096\pm0.017$&$0.326\pm0.001$&$-8.168\pm0.047$&$0.076\pm0.001$& 7522\\
$z$&$2.25\leq\log\sigma_0<2.45$&$1.116\pm0.020$&$0.339\pm0.002$&$-8.455\pm0.055$&$0.072\pm0.001$& 5595\\
$z$&$2.30\leq\log\sigma_0<2.50$&$1.159\pm0.026$&$0.348\pm0.002$&$-8.737\pm0.072$&$0.069\pm0.001$& 3603\\
$z$&$\log\sigma_0\geq2.30$&$1.162\pm0.025$&$0.349\pm0.002$&$-8.748\pm0.070$&$0.069\pm0.001$& 3642\\
$z$&$\log\sigma_0\geq2.35$&$1.230\pm0.041$&$0.355\pm0.002$&$-9.033\pm0.111$&$0.066\pm0.001$& 1869\\
\enddata
\tablecomments{$a$, $b$, and $c$ are the coefficients of the FP (Equation \ref{eq:fp}), while $\varepsilon$ is the intrinsic scatter around the FP (see Equation \ref{eq:chi}). $N$ is the number of ETGs in each category. The units of $\sigma_0$ are km s$^{-1}$.
}
\end{deluxetable*}

\begin{deluxetable*}{crccccc}
\tablecaption{Coefficients of the FP and Intrinsic Scatter around the FP for Different $\mu_e$ Bins\label{tb:fpmu}}
\tabletypesize{\scriptsize}
\tablehead{\colhead{Band} & \colhead{Category} & \colhead{$a$} & \colhead{$b$} & \colhead{$c$} & \colhead{$\varepsilon$} & \colhead{$N$}
}
\startdata
$g$&$\mu_e<19.5$&$0.456\pm0.023$&$0.252\pm0.007$&$-5.778\pm0.142$&$0.104\pm0.002$& 1810\\
$g$&$19.0\leq\mu_e<20.0$&$0.734\pm0.014$&$0.288\pm0.006$&$-7.082\pm0.115$&$0.094\pm0.001$& 4200\\
$g$&$19.5\leq\mu_e<20.5$&$0.912\pm0.009$&$0.309\pm0.004$&$-7.905\pm0.079$&$0.086\pm0.001$& 7300\\
$g$&$20.0\leq\mu_e<21.0$&$1.016\pm0.008$&$0.306\pm0.003$&$-8.068\pm0.069$&$0.080\pm0.001$& 8326\\
$g$&$20.5\leq\mu_e<21.5$&$1.089\pm0.009$&$0.303\pm0.004$&$-8.172\pm0.085$&$0.077\pm0.001$& 6030\\
$g$&$\mu_e\geq21.0$&$1.169\pm0.013$&$0.265\pm0.004$&$-7.545\pm0.090$&$0.080\pm0.001$& 3242\\
\hline
$r$&$\mu_e<19.0$&$0.700\pm0.015$&$0.278\pm0.004$&$-6.598\pm0.086$&$0.093\pm0.001$& 3757\\
$r$&$18.5\leq\mu_e<19.5$&$0.911\pm0.010$&$0.303\pm0.004$&$-7.547\pm0.077$&$0.085\pm0.001$& 6787\\
$r$&$19.0\leq\mu_e<20.0$&$1.022\pm0.008$&$0.307\pm0.003$&$-7.862\pm0.064$&$0.079\pm0.001$& 8652\\
$r$&$19.5\leq\mu_e<20.5$&$1.110\pm0.008$&$0.300\pm0.004$&$-7.922\pm0.075$&$0.076\pm0.001$& 6858\\
$r$&$20.0\leq\mu_e<21.0$&$1.198\pm0.012$&$0.283\pm0.005$&$-7.780\pm0.114$&$0.078\pm0.001$& 3529\\
$r$&$\mu_e\geq20.5$&$1.293\pm0.021$&$0.255\pm0.007$&$-7.408\pm0.165$&$0.082\pm0.002$& 1342\\
\hline
$i$&$\mu_e<18.5$&$0.732\pm0.016$&$0.291\pm0.005$&$-6.807\pm0.096$&$0.091\pm0.001$& 3080\\
$i$&$18.0\leq\mu_e<19.0$&$0.938\pm0.010$&$0.312\pm0.004$&$-7.649\pm0.079$&$0.084\pm0.001$& 6188\\
$i$&$18.5\leq\mu_e<19.5$&$1.053\pm0.008$&$0.312\pm0.003$&$-7.907\pm0.062$&$0.078\pm0.001$& 8610\\
$i$&$19.0\leq\mu_e<20.0$&$1.148\pm0.008$&$0.311\pm0.003$&$-8.113\pm0.067$&$0.074\pm0.001$& 7517\\
$i$&$19.5\leq\mu_e<20.5$&$1.233\pm0.011$&$0.289\pm0.005$&$-7.868\pm0.099$&$0.075\pm0.001$& 4162\\
$i$&$\mu_e\geq20.0$&$1.323\pm0.019$&$0.266\pm0.007$&$-7.588\pm0.148$&$0.081\pm0.002$& 1617\\
\hline
$z$&$\mu_e<18.0$&$0.732\pm0.018$&$0.315\pm0.006$&$-7.145\pm0.110$&$0.088\pm0.001$& 2581\\
$z$&$17.5\leq\mu_e<18.5$&$0.938\pm0.011$&$0.326\pm0.004$&$-7.797\pm0.081$&$0.080\pm0.001$& 5766\\
$z$&$18.0\leq\mu_e<19.0$&$1.056\pm0.008$&$0.329\pm0.003$&$-8.127\pm0.060$&$0.073\pm0.001$& 8594\\
$z$&$18.5\leq\mu_e<19.5$&$1.156\pm0.008$&$0.328\pm0.003$&$-8.340\pm0.062$&$0.069\pm0.001$& 7802\\
$z$&$19.0\leq\mu_e<20.0$&$1.257\pm0.010$&$0.322\pm0.004$&$-8.461\pm0.091$&$0.070\pm0.001$& 4562\\
$z$&$\mu_e\geq19.5$&$1.350\pm0.017$&$0.297\pm0.006$&$-8.172\pm0.126$&$0.075\pm0.002$& 1975\\
\enddata
\tablecomments{$a$, $b$, and $c$ are the coefficients of the FP (Equation \ref{eq:fp}), while $\varepsilon$ is the intrinsic scatter around the FP (see Equation \ref{eq:chi}). $N$ is the number of ETGs in each category. The units of $\mu_e$ are mag arcsec$^{-2}$.
}
\end{deluxetable*}


\begin{deluxetable*}{crcccc}
\tablecaption{Coefficients of the $L$--$\sigma_0$ Relation and Intrinsic Scatter around the Relation \label{tb:ls}}
\tabletypesize{\scriptsize}
\tablehead{\colhead{Band} & \colhead{Category} & \colhead{$p$} & \colhead{$q$} & \colhead{$\varepsilon$} & \colhead{$N$}
}
\startdata
$g$&All&$ 5.83\pm 0.03$&$1.95\pm0.02$&$0.225\pm0.001$&16,779\\
$g$&$\mu_e<19.5$&$ 7.35\pm 0.11$&$1.16\pm0.05$&$0.236\pm0.004$& 1961\\
$g$&$19.5\leq\mu_e<20.0$&$ 6.08\pm 0.07$&$1.78\pm0.03$&$0.193\pm0.003$& 3035\\
$g$&$20.0\leq\mu_e<20.5$&$ 5.60\pm 0.05$&$2.04\pm0.02$&$0.176\pm0.002$& 4497\\
$g$&$20.5\leq\mu_e<21.0$&$ 5.40\pm 0.05$&$2.18\pm0.02$&$0.164\pm0.002$& 3989\\
$g$&$\mu_e\geq21.0$&$ 5.12\pm 0.06$&$2.36\pm0.03$&$0.177\pm0.003$& 3297\\
\hline
$r$&All&$ 5.82\pm 0.03$&$2.01\pm0.02$&$0.222\pm0.001$&16,779\\
$r$&$\mu_e<19.0$&$ 6.69\pm 0.07$&$1.53\pm0.03$&$0.212\pm0.003$& 4001\\
$r$&$19.0\leq\mu_e<19.5$&$ 5.69\pm 0.05$&$2.05\pm0.02$&$0.174\pm0.002$& 4444\\
$r$&$19.5\leq\mu_e<20.0$&$ 5.45\pm 0.05$&$2.20\pm0.02$&$0.162\pm0.002$& 4394\\
$r$&$20.0\leq\mu_e<20.5$&$ 5.12\pm 0.06$&$2.39\pm0.03$&$0.162\pm0.002$& 2572\\
$r$&$\mu_e\geq20.5$&$ 4.71\pm 0.10$&$2.62\pm0.04$&$0.175\pm0.004$& 1368\\
\hline
$i$&All&$ 5.77\pm 0.03$&$2.07\pm0.02$&$0.222\pm0.001$&16,779\\
$i$&$\mu_e<18.5$&$ 6.64\pm 0.08$&$1.59\pm0.03$&$0.208\pm0.003$& 3300\\
$i$&$18.5\leq\mu_e<19.0$&$ 5.68\pm 0.05$&$2.08\pm0.02$&$0.175\pm0.002$& 4189\\
$i$&$19.0\leq\mu_e<19.5$&$ 5.38\pm 0.05$&$2.27\pm0.02$&$0.160\pm0.002$& 4619\\
$i$&$19.5\leq\mu_e<20.0$&$ 5.05\pm 0.05$&$2.46\pm0.02$&$0.154\pm0.002$& 3019\\
$i$&$\mu_e\geq20.0$&$ 4.66\pm 0.09$&$2.69\pm0.04$&$0.175\pm0.004$& 1652\\
\hline
$z$&All&$ 5.94\pm 0.03$&$2.04\pm0.02$&$0.222\pm0.001$&16,779\\
$z$&$\mu_e<18.0$&$ 6.70\pm 0.08$&$1.59\pm0.04$&$0.213\pm0.003$& 2772\\
$z$&$18.0\leq\mu_e<18.5$&$ 5.69\pm 0.06$&$2.11\pm0.03$&$0.170\pm0.002$& 4044\\
$z$&$18.5\leq\mu_e<19.0$&$ 5.38\pm 0.04$&$2.30\pm0.02$&$0.154\pm0.002$& 4764\\
$z$&$19.0\leq\mu_e<19.5$&$ 5.00\pm 0.05$&$2.52\pm0.02$&$0.153\pm0.002$& 3179\\
$z$&$\mu_e\geq19.5$&$ 4.62\pm 0.08$&$2.75\pm0.04$&$0.176\pm0.003$& 2020\\
\enddata
\tablecomments{$p$ and $q$ are the coefficients of the $L$--$\sigma_0$ relation (Equation \ref{eq:ls}), while $\varepsilon$ is the intrinsic scatter around the $L$--$\sigma_0$ relation (see Equation \ref{eq:chi2}). $N$ is the number of ETGs in each category. We note that ETGs with $M_r>-19.5$ were included to fit the $L$--$\sigma_0$ relation. The units of $\mu_e$ are mag arcsec$^{-2}$.
}
\end{deluxetable*}


\begin{deluxetable*}{crcccc}
\tablecaption{Coefficients of the $R_e$--$\mu_e$ Relation and Intrinsic Scatter around the Relation \label{tb:rm}}
\tabletypesize{\scriptsize}
\tablehead{\colhead{Band} & \colhead{Category} & \colhead{$\alpha$} & \colhead{$\beta$} & \colhead{$\varepsilon$} & \colhead{$N$}
}
\startdata
$g$&All&$-5.08\pm 0.03$&$0.269\pm0.001$&$0.139\pm0.001$&16,283\\
$g$&$2.00\leq\log\sigma_0<2.20$&$-4.85\pm 0.03$&$0.253\pm0.001$&$0.101\pm0.001$& 8010\\
$g$&$2.05\leq\log\sigma_0<2.25$&$-5.10\pm 0.03$&$0.268\pm0.001$&$0.103\pm0.001$& 8951\\
$g$&$2.10\leq\log\sigma_0<2.30$&$-5.33\pm 0.03$&$0.281\pm0.001$&$0.105\pm0.001$& 9324\\
$g$&$2.15\leq\log\sigma_0<2.35$&$-5.61\pm 0.03$&$0.297\pm0.002$&$0.103\pm0.001$& 8805\\
$g$&$2.20\leq\log\sigma_0<2.40$&$-5.79\pm 0.03$&$0.307\pm0.002$&$0.102\pm0.001$& 7522\\
$g$&$2.25\leq\log\sigma_0<2.45$&$-5.97\pm 0.04$&$0.318\pm0.002$&$0.099\pm0.001$& 5595\\
$g$&$2.30\leq\log\sigma_0<2.50$&$-6.17\pm 0.05$&$0.330\pm0.002$&$0.096\pm0.001$& 3603\\
$g$&$\log\sigma_0\geq2.30$&$-6.15\pm 0.05$&$0.329\pm0.002$&$0.097\pm0.001$& 3642\\
$g$&$\log\sigma_0\geq2.35$&$-6.33\pm 0.06$&$0.339\pm0.003$&$0.092\pm0.002$& 1869\\
\hline
$r$&All&$-4.87\pm 0.03$&$0.269\pm0.002$&$0.141\pm0.001$&16,283\\
$r$&$2.00\leq\log\sigma_0<2.20$&$-4.67\pm 0.03$&$0.254\pm0.002$&$0.101\pm0.001$& 8010\\
$r$&$2.05\leq\log\sigma_0<2.25$&$-4.92\pm 0.03$&$0.269\pm0.002$&$0.103\pm0.001$& 8951\\
$r$&$2.10\leq\log\sigma_0<2.30$&$-5.17\pm 0.03$&$0.284\pm0.002$&$0.105\pm0.001$& 9324\\
$r$&$2.15\leq\log\sigma_0<2.35$&$-5.45\pm 0.03$&$0.301\pm0.002$&$0.102\pm0.001$& 8805\\
$r$&$2.20\leq\log\sigma_0<2.40$&$-5.66\pm 0.03$&$0.313\pm0.002$&$0.100\pm0.001$& 7522\\
$r$&$2.25\leq\log\sigma_0<2.45$&$-5.83\pm 0.04$&$0.324\pm0.002$&$0.097\pm0.001$& 5595\\
$r$&$2.30\leq\log\sigma_0<2.50$&$-6.01\pm 0.04$&$0.335\pm0.002$&$0.093\pm0.001$& 3603\\
$r$&$\log\sigma_0\geq2.30$&$-5.99\pm 0.04$&$0.334\pm0.002$&$0.094\pm0.001$& 3642\\
$r$&$\log\sigma_0\geq2.35$&$-6.12\pm 0.06$&$0.343\pm0.003$&$0.089\pm0.002$& 1869\\
\hline
$i$&All&$-4.76\pm 0.03$&$0.268\pm0.002$&$0.144\pm0.001$&16,283\\
$i$&$2.00\leq\log\sigma_0<2.20$&$-4.64\pm 0.03$&$0.257\pm0.002$&$0.100\pm0.001$& 8010\\
$i$&$2.05\leq\log\sigma_0<2.25$&$-4.90\pm 0.03$&$0.273\pm0.002$&$0.103\pm0.001$& 8951\\
$i$&$2.10\leq\log\sigma_0<2.30$&$-5.14\pm 0.03$&$0.288\pm0.002$&$0.104\pm0.001$& 9324\\
$i$&$2.15\leq\log\sigma_0<2.35$&$-5.42\pm 0.03$&$0.305\pm0.002$&$0.102\pm0.001$& 8805\\
$i$&$2.20\leq\log\sigma_0<2.40$&$-5.63\pm 0.03$&$0.319\pm0.002$&$0.100\pm0.001$& 7522\\
$i$&$2.25\leq\log\sigma_0<2.45$&$-5.80\pm 0.04$&$0.330\pm0.002$&$0.096\pm0.001$& 5595\\
$i$&$2.30\leq\log\sigma_0<2.50$&$-5.98\pm 0.04$&$0.341\pm0.002$&$0.091\pm0.001$& 3603\\
$i$&$\log\sigma_0\geq2.30$&$-5.97\pm 0.04$&$0.340\pm0.002$&$0.093\pm0.001$& 3642\\
$i$&$\log\sigma_0\geq2.35$&$-6.10\pm 0.06$&$0.349\pm0.003$&$0.086\pm0.002$& 1869\\
\hline
$z$&All&$-4.58\pm 0.03$&$0.263\pm0.002$&$0.143\pm0.001$&16,283\\
$z$&$2.00\leq\log\sigma_0<2.20$&$-4.65\pm 0.03$&$0.262\pm0.002$&$0.097\pm0.001$& 8010\\
$z$&$2.05\leq\log\sigma_0<2.25$&$-4.88\pm 0.03$&$0.276\pm0.002$&$0.101\pm0.001$& 8951\\
$z$&$2.10\leq\log\sigma_0<2.30$&$-5.11\pm 0.03$&$0.291\pm0.002$&$0.102\pm0.001$& 9324\\
$z$&$2.15\leq\log\sigma_0<2.35$&$-5.39\pm 0.03$&$0.309\pm0.002$&$0.099\pm0.001$& 8805\\
$z$&$2.20\leq\log\sigma_0<2.40$&$-5.61\pm 0.03$&$0.323\pm0.002$&$0.096\pm0.001$& 7522\\
$z$&$2.25\leq\log\sigma_0<2.45$&$-5.76\pm 0.04$&$0.333\pm0.002$&$0.092\pm0.001$& 5595\\
$z$&$2.30\leq\log\sigma_0<2.50$&$-5.92\pm 0.04$&$0.344\pm0.002$&$0.087\pm0.001$& 3603\\
$z$&$\log\sigma_0\geq2.30$&$-5.90\pm 0.04$&$0.343\pm0.002$&$0.088\pm0.001$& 3642\\
$z$&$\log\sigma_0\geq2.35$&$-6.02\pm 0.05$&$0.352\pm0.003$&$0.081\pm0.001$& 1869\\
\enddata
\tablecomments{$\alpha$ and $\beta$ are the coefficients of the $R_e$--$\mu_e$ relation (Equation \ref{eq:rm}), while $\varepsilon$ is the intrinsic scatter around the $R_e$--$\mu_e$ relation (see Equation \ref{eq:chi2}). $N$ is the number of ETGs in each category. The units of $\sigma_0$ are km s$^{-1}$.
}
\end{deluxetable*}


\clearpage

\begin{thebibliography}{}
\bibitem[Abazajian et al.(2009)]{Abazajian2009} Abazajian, K.~N., Adelman-McCarthy, J.~K., Ag{\"u}eros, M.~A., et al.\ 2009, \apjs, 182, 543. doi:10.1088/0067-0049/182/2/543
\bibitem[Aguado et al.(2019)]{Aguado2019} Aguado, D.~S., Ahumada, R., Almeida, A., et al.\ 2019, \apjs, 240, 23. doi:10.3847/1538-4365/aaf651
\bibitem[Bernardi et al.(2020)]{Bernardi2020} Bernardi, M., Dom{\'\i}nguez S{\'a}nchez, H., Margalef-Bentabol, B., et al.\ 2020, \mnras, 494, 5148. doi:10.1093/mnras/staa1064
\bibitem[Bernardi et al.(2007)]{Bernardi2007} Bernardi, M., Hyde, J.~B., Sheth, R.~K., et al.\ 2007, \aj, 133, 1741. doi:10.1086/511783
\bibitem[Bernardi et al.(2011a)]{Bernardi2011a} Bernardi, M., Roche, N., Shankar, F., et al.\ 2011a, \mnras, 412, 684. doi:10.1111/j.1365-2966.2010.17984.x
\bibitem[Bernardi et al.(2011b)]{Bernardi2011b} Bernardi, M., Roche, N., Shankar, F., et al.\ 2011b, \mnras, 412, L6. doi:10.1111/j.1745-3933.2010.00982.x
\bibitem[Bernardi et al.(2003a)]{Bernardi2003a} Bernardi, M., Sheth, R.~K., Annis, J., et al.\ 2003a, \aj, 125, 1817. doi:10.1086/367776
\bibitem[Bernardi et al.(2003b)]{Bernardi2003b} Bernardi, M., Sheth, R.~K., Annis, J., et al.\ 2003b, \aj, 125, 1866. doi:10.1086/367794
\bibitem[Bertin et al.(2002)]{Bertin2002} Bertin, G., Ciotti, L., \& Del Principe, M.\ 2002, \aap, 386, 149. doi:10.1051/0004-6361:20020248
\bibitem[Blanton \& Roweis(2007)]{Blanton2007} Blanton, M.~R. \& Roweis, S.\ 2007, \aj, 133, 734. doi:10.1086/510127
\bibitem[Bruzual \& Charlot(2003)]{Bruzual2003} Bruzual, G. \& Charlot, S.\ 2003, \mnras, 344, 1000. doi:10.1046/j.1365-8711.2003.06897.x
\bibitem[Cappellari et al.(2006)]{Cappellari2006} Cappellari, M., Bacon, R., Bureau, M., et al.\ 2006, \mnras, 366, 1126. doi:10.1111/j.1365-2966.2005.09981.x
\bibitem[Cappellari et al.(2013)]{Cappellari2013} Cappellari, M., Scott, N., Alatalo, K., et al.\ 2013, \mnras, 432, 1709. doi:10.1093/mnras/stt562
\bibitem[Chabrier(2003)]{Chabrier2003} Chabrier, G.\ 2003, \pasp, 115, 763. doi:10.1086/376392
\bibitem[Choi et al.(2007)]{Choi2007} Choi, Y.-Y., Park, C., \& Vogeley, M.~S.\ 2007, \apj, 658, 884. doi:10.1086/511060
\bibitem[Choi et al.(2010)]{Choi2010} Choi, Y.-Y., Han, D.-H., \& Kim, S.~S.\ 2010, Journal of Korean Astronomical Society, 43, 191. doi:10.5303/JKAS.2010.43.6.191
\bibitem[de Graaff et al.(2020)]{deGraaff2020} de Graaff, A., Bezanson, R., Franx, M., et al.\ 2020, \apjl, 903, L30. doi:10.3847/2041-8213/abc428
\bibitem[de Graaff et al.(2021)]{deGraaff2021} de Graaff, A., Bezanson, R., Franx, M., et al.\ 2021, \apj, 913, 103. doi:10.3847/1538-4357/abf1e7
\bibitem[Dekel \& Cox(2006)]{Dekel2006} Dekel, A. \& Cox, T.~J.\ 2006, \mnras, 370, 1445. doi:10.1111/j.1365-2966.2006.10566.x
\bibitem[Desroches et al.(2007)]{Desroches2007} Desroches, L.-B., Quataert, E., Ma, C.-P., et al.\ 2007, \mnras, 377, 402. doi:10.1111/j.1365-2966.2007.11612.x
\bibitem[Djorgovski \& Davis(1987)]{Djorgovski1987} Djorgovski, S. \& Davis, M.\ 1987, \apj, 313, 59. doi:10.1086/164948
\bibitem[D'Onofrio et al.(2008)]{D'Onofrio2008} D'Onofrio, M., Fasano, G., Varela, J., et al.\ 2008, \apj, 685, 875. doi:10.1086/591143
\bibitem[Dressler et al.(1987)]{Dressler1987} Dressler, A., Lynden-Bell, D., Burstein, D., et al.\ 1987, \apj, 313, 42. doi:10.1086/164947
\bibitem[Faber \& Jackson(1976)]{Faber1976} Faber, S.~M. \& Jackson, R.~E.\ 1976, \apj, 204, 668. doi:10.1086/154215
\bibitem[Gallazzi et al.(2006)]{Gallazzi2006} Gallazzi, A., Charlot, S., Brinchmann, J., et al.\ 2006, \mnras, 370, 1106. doi:10.1111/j.1365-2966.2006.10548.x
\bibitem[Gargiulo et al.(2009)]{Gargiulo2009} Gargiulo, A., Haines, C.~P., Merluzzi, P., et al.\ 2009, \mnras, 397, 75. doi:10.1111/j.1365-2966.2009.14801.x
\bibitem[Graham et al.(2018)]{Graham2018} Graham, M.~T., Cappellari, M., Li, H., et al.\ 2018, \mnras, 477, 4711. doi:10.1093/mnras/sty504
\bibitem[Graves et al.(2009a)]{Graves2009a} Graves, G.~J., Faber, S.~M., \& Schiavon, R.~P.\ 2009a, \apj, 693, 486. doi:10.1088/0004-637X/693/1/486
\bibitem[Graves et al.(2009b)]{Graves2009b} Graves, G.~J., Faber, S.~M., \& Schiavon, R.~P.\ 2009b, \apj, 698, 1590. doi:10.1088/0004-637X/698/2/1590
\bibitem[Hilz et al.(2012)]{Hilz2012} Hilz, M., Naab, T., Ostriker, J.~P., et al.\ 2012, \mnras, 425, 3119. doi:10.1111/j.1365-2966.2012.21541.x
\bibitem[Hopkins et al.(2008)]{Hopkins2008} Hopkins, P.~F., Cox, T.~J., \& Hernquist, L.\ 2008, \apj, 689, 17. doi:10.1086/592105
\bibitem[Hopkins et al.(2009)]{Hopkins2009} Hopkins, P.~F., Hernquist, L., Cox, T.~J., et al.\ 2009, \apj, 691, 1424. doi:10.1088/0004-637X/691/2/1424
\bibitem[Hyde \& Bernardi(2009a)]{Hyde2009a} Hyde, J.~B. \& Bernardi, M.\ 2009a, \mnras, 394, 1978. doi:10.1111/j.1365-2966.2009.14445.x
\bibitem[Hyde \& Bernardi(2009b)]{Hyde2009b} Hyde, J.~B. \& Bernardi, M.\ 2009b, \mnras, 396, 1171. doi:10.1111/j.1365-2966.2009.14783.x
\bibitem[Jorgensen et al.(1995)]{Jorgensen1995} Jorgensen, I., Franx, M., \& Kjaergaard, P.\ 1995, \mnras, 276, 1341. doi:10.1093/mnras/276.4.1341
\bibitem[Jorgensen et al.(1996)]{Jorgensen1996} Jorgensen, I., Franx, M., \& Kjaergaard, P.\ 1996, \mnras, 280, 167. doi:10.1093/mnras/280.1.167
\bibitem[Jun \& Im(2008)]{Jun2008} Jun, H.~D. \& Im, M.\ 2008, \apjl, 678, L97. doi:10.1086/588552
\bibitem[Kormendy(1977)]{Kormendy1977} Kormendy, J.\ 1977, \apj, 218, 333. doi:10.1086/155687
\bibitem[Kormendy \& Bender(2013)]{Kormendy2013} Kormendy, J. \& Bender, R.\ 2013, \apjl, 769, L5. doi:10.1088/2041-8205/769/1/L5
\bibitem[Kormendy et al.(2009)]{Kormendy2009} Kormendy, J., Fisher, D.~B., Cornell, M.~E., et al.\ 2009, \apjs, 182, 216. doi:10.1088/0067-0049/182/1/216
\bibitem[La Barbera et al.(2010)]{LaBarbera2010} La Barbera, F., de Carvalho, R.~R., de La Rosa, I.~G., et al.\ 2010, \mnras, 408, 1335. doi:10.1111/j.1365-2966.2010.17091.x
\bibitem[Lauer et al.(2007)]{Lauer2007} Lauer, T.~R., Faber, S.~M., Richstone, D., et al.\ 2007, \apj, 662, 808. doi:10.1086/518223
\bibitem[L{\'o}pez-Sanjuan et al.(2012)]{Lopez2012} L{\'o}pez-Sanjuan, C., Le F{\`e}vre, O., Ilbert, O., et al.\ 2012, \aap, 548, A7. doi:10.1051/0004-6361/201219085
\bibitem[Mihos \& Hernquist(1994)]{Mihos1994} Mihos, J.~C. \& Hernquist, L.\ 1994, \apjl, 437, L47. doi:10.1086/187679
\bibitem[Nair \& Abraham(2010)]{Nair2010} Nair, P.~B. \& Abraham, R.~G.\ 2010, \apjs, 186, 427. doi:10.1088/0067-0049/186/2/427
\bibitem[Nigoche-Netro et al.(2009)]{Nigoche-Netro2009} Nigoche-Netro, A., Ruelas-Mayorga, A., \& Franco-Balderas, A.\ 2009, \mnras, 392, 1060. doi:10.1111/j.1365-2966.2008.14145.x
\bibitem[Nipoti et al.(2003)]{Nipoti2003} Nipoti, C., Londrillo, P., \& Ciotti, L.\ 2003, \mnras, 342, 501. doi:10.1046/j.1365-8711.2003.06554.x
\bibitem[O'Leary et al.(2021)]{OLeary2021} O'Leary, J.~A., Moster, B.~P., Naab, T., et al.\ 2021, \mnras, 501, 3215. doi:10.1093/mnras/staa3746
\bibitem[Oogi \& Habe(2013)]{Oogi2013} Oogi, T. \& Habe, A.\ 2013, \mnras, 428, 641. doi:10.1093/mnras/sts047
\bibitem[Oogi et al.(2016)]{Oogi2016} Oogi, T., Habe, A., \& Ishiyama, T.\ 2016, \mnras, 456, 300. doi:10.1093/mnras/stv2581
\bibitem[Park \& Choi(2005)]{PC2005} Park, C. \& Choi, Y.-Y.\ 2005, \apjl, 635, L29. doi:10.1086/499243
\bibitem[Robertson et al.(2006)]{Robertson2006} Robertson, B., Cox, T.~J., Hernquist, L., et al.\ 2006, \apj, 641, 21. doi:10.1086/500360
\bibitem[Saglia et al.(2001)]{Saglia2001} Saglia, R.~P., Colless, M., Burstein, D., et al.\ 2001, \mnras, 324, 389. doi:10.1046/j.1365-8711.2001.04317.x
\bibitem[Samir et al.(2020)]{Samir2020} Samir, R.~M., Takey, A., \& Shaker, A.~A.\ 2020, \apss, 365, 142. doi:10.1007/s10509-020-03857-8
\bibitem[Saulder et al.(2013)]{Saulder2013} Saulder, C., Mieske, S., Zeilinger, W.~W., et al.\ 2013, \aap, 557, A21. doi:10.1051/0004-6361/201321466
\bibitem[Schawinski et al.(2014)]{Schawinski2014} Schawinski, K., Urry, C.~M., Simmons, B.~D., et al.\ 2014, \mnras, 440, 889. doi:10.1093/mnras/stu327
\bibitem[Schlegel et al.(1998)]{Schlegel1998} Schlegel, D.~J., Finkbeiner, D.~P., \& Davis, M.\ 1998, \apj, 500, 525. doi:10.1086/305772
\bibitem[Trujillo et al.(2004)]{Trujillo2004} Trujillo, I., Burkert, A., \& Bell, E.~F.\ 2004, \apjl, 600, L39. doi:10.1086/381528
\bibitem[Willmer(2018)]{Wilmer2018} Willmer, C.~N.~A.\ 2018, \apjs, 236, 47. doi:10.3847/1538-4365/aabfdf
\bibitem[Yoon et al.(2017)]{Yoon2017} Yoon, Y., Im, M., \& Kim, J.-W.\ 2017, \apj, 834, 73. doi:10.3847/1538-4357/834/1/73
\bibitem[Yoon \& Lim(2020)]{YL2020} Yoon, Y. \& Lim, G.\ 2020, \apj, 905, 154. doi:10.3847/1538-4357/abc621
\bibitem[Yoon \& Park(2020)]{YP2020} Yoon, Y. \& Park, C.\ 2020, \apj, 897, 121. doi:10.3847/1538-4357/ab9b26
\bibitem[Yoon et al.(2022)]{Yoon2022} Yoon, Y., Park, C., Chung, H., et al.\ 2022, \apj, 925, 168. doi:10.3847/1538-4357/ac415d
\bibitem[Zaritsky et al.(2006)]{Zaritsky2006} Zaritsky, D., Gonzalez, A.~H., \& Zabludoff, A.~I.\ 2006, \apj, 638, 725. doi:10.1086/498672
\end{thebibliography}
\end{document}